%% file: ttHmultilep.tex
\documentclass[a4paper,UKenglish,5p,times]{elsarticle}
\journal{Physics Letters B}
\pdfoutput=1
\biboptions{sort&compress}
\usepackage[hyperindex,breaklinks]{hyperref}

\usepackage[minimal]{atlaspackage}
\usepackage[texmf]{atlasphysics}
\usepackage{epstopdf}
\usepackage{subfig}

\usepackage{multicol}
\usepackage{multirow}

\usepackage{preprintcover}

\PreprintJournalName{Phys.\ Lett.\ B}
\PreprintIdNumber{CERN-PH-EP-2015-109}

\input{Macros.tex}

\date{\today} 

\newcommand{\AtlasTitle}{Search for the associated production of the Higgs boson with a top quark
 pair in multilepton final states with the ATLAS detector}
\newcommand{\AtlasAbstract}{A search for the associated production of the Higgs boson with a top quark pair is performed in multilepton final states using 20.3~fb$^{-1}$ of proton--proton collision data recorded by the ATLAS experiment at $\sqrt{s}=8$~TeV at the Large Hadron Collider. Five final states, targeting the decays $H\to WW^*$, $\tau\tau$, and $ZZ^*$, are examined for the presence of the Standard Model (SM) Higgs boson: two same-charge light leptons ($e$ or $\mu$) without a hadronically decaying $\tau$ lepton; three light leptons; two same-charge light leptons with a hadronically decaying $\tau$ lepton; four light leptons; and one light lepton and two hadronically decaying $\tau$ leptons. No significant excess of events is observed above the background expectation. The best fit for the \tth{} production cross section, assuming a Higgs boson mass of 125~GeV, is $2.1 ^{+1.4}_{-1.2}$ times the SM expectation, and the observed (expected) upper limit at the 95\% confidence level is 4.7 (2.4) times the SM rate.  The $p$-value for compatibility with the background-only hypothesis is $1.8\sigma$; the expectation in the presence of a Standard Model signal is $0.9\sigma$.}

\PreprintCoverPaperTitle{\AtlasTitle}

\PreprintCoverAbstract{\AtlasAbstract}

\begin{document}

\title{\AtlasTitle}
\begin{abstract}
\AtlasAbstract
\end{abstract}
\author{The ATLAS Collaboration}
\maketitle
\sloppy
\input{Intro.tex}
\input{Detector.tex}
\input{Simulation.tex}

\input{Objects.tex}

\input{Selection.tex} 
\input{Background.tex}

\input{Syst.tex}

\input{Results.tex}

\input{Conclusions.tex}

\section*{Acknowledgements}
\input{Acknowledgements.tex}

\section*{References}
\bibliographystyle{atlasBibStyleWithTitle}
\bibliography{ttHmultilep}

\onecolumn\newpage \input{atlas_authlist}
\end{document}

%% file: Macros.tex
\newcommand{\tth}{\ensuremath{t\bar{t}H}\xspace}
\newcommand{\ttv}{\ensuremath{t\bar{t}V}\xspace}
\newcommand{\ttw}{\ensuremath{t\bar{t}\Wboson}\xspace}
\newcommand{\ttwp}{\ensuremath{t\bar{t}\Wboson^{+}}\xspace}
\newcommand{\ttwn}{\ensuremath{t\bar{t}\Wboson^{-}}\xspace}

\newcommand{\ttz}{\ensuremath{t\bar{t}\Zboson}\xspace}

\newcommand{\tz}{\ensuremath{t\Zboson}\xspace}
\newcommand{\ttj}{\ensuremath{t\bar{t}+}jets\xspace}
\newcommand{\zj}{\ensuremath{\Zboson \rightarrow \ell^+\ell^-+}jets\xspace}
\newcommand{\wj}{\ensuremath{\Wboson \rightarrow \ell\nu+}jets\xspace}
\newcommand{\zdyj}{\ensuremath{\Zboson/\gamma^{*} \rightarrow \ell^+\ell^-+}jets\xspace}

\newcommand{\tauh}{\ensuremath{\tau_\mathrm{had}}\xspace}

\newcommand{\oneltwotau}{\ensuremath{1\ell2\tauh}\xspace}
\newcommand{\twolnotau}{\ensuremath{2\ell0\tauh}\xspace}
\newcommand{\twoltau}{\ensuremath{2\ell1\tauh}\xspace}
\newcommand{\threel}{\ensuremath{3\ell}\xspace}
\newcommand{\fourl}{\ensuremath{4\ell}\xspace}
\newcommand{\njet}{\ensuremath{n_\mathrm{jet}}\xspace}

\newcommand{\obsjust}[1]{\makebox[0.4cm]{\null\hfill{}#1}}

%% file: Intro.tex
\section{Introduction} \label{sec:intro}

The discovery of a new particle $H$ with a mass of about 125 GeV in searches for the Standard Model (SM) \cite{Glashow:1961tr,Weinberg:1967tq,sm_salam} Higgs boson \cite{Englert:1964et,higgs12,Higgs:1964pj,ghk} at the LHC was reported by the ATLAS \cite{2012gk} and CMS \cite{2012gu} 
Collaborations in July 2012. 
The particle has been observed in the decays $H\rightarrow \gamma \gamma$ \cite{Aad:2014eha,Khachatryan:2014ira}, $H\rightarrow ZZ^* \rightarrow 4\ell$ \cite{Aad:2014eva,Chatrchyan:2013mxa}, and
$H\rightarrow WW^* \rightarrow \ell\nu\ell\nu$ \cite{ATLAS:2014aga,Chatrchyan:2013iaa}, and evidence has been reported for $H\rightarrow \tau\tau$ \cite{Aad:2015vsa,Chatrchyan:2014nva}, consistent with the rates expected for the SM Higgs boson. 

The observation of the process in which the Higgs boson is produced in association with a pair of top quarks (\tth) would permit a direct measurement of the top quark--Higgs boson Yukawa coupling in a process that is tree-level at the lowest order, which is otherwise accessible primarily through loop effects.  Having both the tree- and loop-level measurements would allow disambiguation of new physics effects that could affect the two differently, such as dimension-six operators contributing to the $ggH$ vertex.  This letter describes a search for the SM Higgs boson in the \tth~production mode in multilepton final states. The five final states considered are: two same-charge-sign light leptons ($e$ or $\mu$) with no additional hadronically decaying $\tau$ lepton; three light leptons; two same-sign light leptons with one hadronically decaying $\tau$ lepton; four light leptons; and one light lepton with two hadronically decaying $\tau$ candidates.  These channels are sensitive to the Higgs decays $H\to WW^*$, $\tau\tau$, and $ZZ^*$ produced in association with a top quark pair decaying to one or two leptons.  A similar search has been performed by the CMS Collaboration~\cite{Khachatryan:2014qaa}. 

The selections of this search are designed to avoid overlap with ATLAS searches for \tth in $H\rightarrow b\bar{b}$ \cite{tthbb} and $H\rightarrow \gamma \gamma$ \cite{2014tthdiph} decays. 
The main backgrounds to the signal arise from $t\bar{t}$ production with additional jets and non-prompt leptons, associated production of a top quark pair and a vector boson $W$ or $Z$ (collectively denoted \ttv), and other processes where the electron charge is incorrectly measured or where quark or gluon jets are incorrectly identified as $\tau$ candidates.

%% file: Detector.tex
\section{ATLAS detector and dataset}

The features of the ATLAS detector \cite{detectorpaper} most relevant to this analysis are briefly summarized here. The detector consists of an inner tracking detector system surrounded by a superconducting solenoid, electromagnetic and hadronic calorimeters, and a muon spectrometer. 
Charged particles in the pseudorapidity\footnote{
  The ATLAS experiment uses a right-handed coordinate system with its origin at the nominal interaction point (IP) in the centre
  of the detector, and the $z$-axis along the beam line. The $x$-axis points from the IP to the centre of the LHC ring, and the
  $y$-axis points upwards. Cylindrical coordinates $(r, \phi)$ are used in the transverse plane, $\phi$ being the azimuthal angle
  around the $z$-axis. Observables labelled ``transverse'' are projected onto the $x$--$y$ plane. The pseudorapidity is defined
  in terms of the polar angle $\theta$ as $\eta=-\ln\tan \theta/2$. The transverse momentum is defined as $\pt=p\sin\theta=p/\cosh\eta$,
  and the transverse energy \et has an analogous definition.}
range $|\eta| < 2.5$ are reconstructed with the inner tracking detector, which is immersed in a 2~$\mathrm{T}$
magnetic field parallel to the detector axis and consists of pixel and strip semiconductor detectors as well as a straw-tube transition radiation tracker.
The solenoid is surrounded by a calorimeter system covering $|\eta| < 4.9$, which provides three-dimensional reconstruction of particle
showers.
Lead/liquid-argon (LAr) sampling technology is used for the electromagnetic component.
Iron/scintillator-tile sampling calorimeters are used for the hadronic component for $|\eta| < 1.7$, and copper/LAr and tungsten/LAr
technology is used for $|\eta| > 1.5$.
Outside the calorimeter system, air-core toroids provide a magnetic field for the muon spectrometer. 
Three stations of precision drift tubes and cathode-strip chambers provide a measurement of the muon track position and curvature in the region
$|\eta| < 2.7$.
Resistive-plate and thin-gap chambers provide muon triggering capability up to $|\eta| = 2.4$.

This search uses data collected by the ATLAS experiment in 2012 at 
a centre-of-mass energy of $\sqrt{s}=8$~TeV.
All events considered were recorded while the detector and trigger systems were fully functional; the integrated luminosity of this dataset is 20.3~\ifb{}.

%% file: Simulation.tex
\section{\label{ssec:xsections}Cross sections for signal and background processes}
The cross section for the production of $\tth$ in $pp$ collisions has been calculated at next-to-leading order (NLO) in quantum chromodynamics (QCD) \cite{Dawson:2003zu,Reina:2001sf,Beenakker:2002nc,Beenakker:2001rj,lhcxs}.  Uncertainties on the cross section are evaluated by varying the renormalization and factorization scales by factors of two and by varying the input parton distribution functions (PDF) of the proton.  A Higgs boson mass of $m_H = 125\gev$ is assumed; this gives a predicted \tth production cross section at $\sqrt{s} = 8\tev$ of 129 $^{+5}_{-12}$ (scale) $\pm$ 10 (PDF) fb \cite{Heinemeyer:2013tqa}.  This assumed Higgs boson mass is consistent with the combined ATLAS and CMS measurement~\cite{Aad:2015zhl}.

In this letter the associated production of single top quarks with a Higgs boson is considered a background process and set to the Standard Model rate.  The production of $tHqb$ and $tHW$ is taken into account.  In the Standard Model these rates are very small compared to \tth production.  These processes are simulated with the same parameters as used by the ATLAS \tth, $H\to\gamma\gamma$ search \cite{2014tthdiph}.  The cross sections for both are computed using the \textsc{MG5\_aMC@NLO} generator \cite{Alwall:2014hca} at NLO in QCD.  For $tHqb$, the renormalization and factorization scales are set to 75 GeV and the process is computed in the four-flavour scheme, yielding $\sigma(tHqb) = 17.2$ $^{+0.8}_{-1.4}$ (scale) $^{+1.2}_{-0.9}$ (PDF) fb. For $tHW$, dynamic factorization and renormalization scales are used, and the process is computed in the five-flavour scheme; the result is $\sigma(tHW) = 4.7$ $^{+0.4}_{-0.3}$ (scale) $^{+0.8}_{-0.6}$ (PDF) fb.  The interference of $tHW$ production with \tth, which appears at NLO for $tHW$ in diagrams with an additional $b$-quark in the final state, is not considered.

The production of $\ttw$ and $\ttbar (Z/\gamma^*) \to \ttbar \ell^+ \ell^-$ yield multilepton final states with $b$-quarks and are major backgrounds to the \tth signal. For simplicity of notation the latter process is referred to as $\ttz$ throughout this letter with off-shell $Z$ and photon components also included except where noted otherwise.  The $\ttw$ process includes both $\ttwp$ and $\ttwn$ components. Next-to-leading-order cross sections are used for \ttw~\cite{Campbell:2012dh} and \ttz~\cite{Garzelli:2012bn}.  The \textsc{MG5\_aMC@NLO} generator is used to reproduce the QCD scale uncertainties of these calculations and determine uncertainties due to the PDF. For \ttw production the value 232 $\pm$ 28 (scale) $\pm$ 18 (PDF) fb is used, and for \ttz production\footnote{The NLO cross section is only evaluated for \ttz production with on-shell \Zboson.  The cross section obtained for $\ttbar (Z/\gamma^*)$ production including off-shell $\Zboson/\gamma^*$ contributions in a leading-order simulation is scaled by a $K$-factor of 1.35  obtained as the ratio of NLO and LO on-shell cross sections.  The $K$-factor differs from that of Ref.~\cite{Garzelli:2012bn} due to a different choice of PDF.} the value is 206 $\pm$ 23 (scale) $\pm$ 18 (PDF) fb.  

The associated production of a single top quark and a $Z$ boson is a subleading background for the most sensitive channels. The cross section has been calculated at NLO for the $t$- and $s$-channels~\cite{Campbell:2013yla}. The resulting values used in this work are 160 $\pm$ 7 (scale) $\pm$ 11 (PDF) fb for $tZ$ and 76 $\pm$ 4 (scale) $\pm$ 5 (PDF) fb for $\bar{t}Z$.
The cross section for the production of $tWZ$ is computed at leading order (LO) using the \textsc{MadGraph} v5 generator \cite{mg5} and found to be 4.1 fb.

The cross section for inclusive production of vector boson pairs $WW$, $WZ$, and $ZZ$ is computed using MCFM \cite{Campbell:2011bn}.  Contributions from virtual photons and off-shell $Z$ bosons are included.  The uncertainties on the acceptance for these processes in the signal regions (which favour production with additional $b$- or $c$-quarks) dominate over the inclusive cross-section uncertainty (see Section~\ref{sec:otherpromptbkg}) and so the latter is neglected in the analysis.

The inclusive \ttbar cross section is calculated at next-to-next-to-leading order (NNLO) in QCD 
which includes resummation of next-to-next-to-leading logarithmic (NNLL) soft gluon terms
using \textsc{Top++}~\cite{Czakon:2011xx}, yielding
$253^{+13}_{-15}$~pb for $\sqrt{s} = 8$ \tev. The single-top-quark samples are normalized to
the approximate NNLO theoretical cross sections~\cite{Kidonakis:2010tc,Kidonakis:2011wy,Kidonakis:2010ux}
using the {\sc MSTW2008} \cite{Martin:2009iq} NNLO PDF set. The production of \zj and \wj is normalized using NNLO cross sections as computed by FEWZ~\cite{Anastasiou:2003ds}.

\section{Event generation}
\begin{table*}
\begin{center}
\caption{\label{tbl:evgen} Configurations used for event generation of signal and background processes.  If only one parton distribution function is shown, the same one is used for both the matrix element (ME) and parton shower generators; if two are shown, the first is used for the matrix element calculation and the second for the parton shower.  ``Tune'' refers to the underlying-event tune of the parton shower generator. ``\textsc{Pythia} 6'' refers to version 6.425; ``\textsc{Pythia} 8'' refers to version 8.1; ``\textsc{Herwig++}'' refers to version 2.6; ``\textsc{MadGraph}'' refers to version 5;  ``\textsc{Alpgen}'' refers to version 2.14; ``\textsc{Sherpa}'' refers to version 1.4; ``\textsc{gg2ZZ}'' refers to version 2.0.}
{\small
\begin{tabular}{lllll}
\hline\hline
Process & ME Generator & Parton Shower & PDF & Tune\\ 
\hline
\tth & \textsc{HELAC}-Oneloop \cite{Bevilacqua:2011xh,Garzelli:2011vp} & \textsc{Pythia} 8 \cite{Pythia8} & CT10 \cite{ct10}/CTEQ6L1 \cite{cteq6l1,cteq6} & AU2 \cite{ATLASUETune0}\\
 & + \textsc{Powheg}-BOX \cite{Nason:2004rx,Frixione:2007vw,Alioli:2010xd} & \\
$tHqb$ & \textsc{MadGraph} \cite{mg5} & \textsc{Pythia} 8 & CT10 & AU2 \\
$tHW$ & \textsc{MG5\_aMC@NLO} \cite{Alwall:2014hca} & \textsc{Herwig++} \cite{Bahr:2008pv} & CT10/MRST LO** \cite{Sherstnev:2007nd} & UE-EE-4 \cite{Gieseke:2012ft} \\
$\ttbar W$ + $\le 2$ partons & \textsc{MadGraph} & \textsc{Pythia} 6 \cite{Pythia6} & CTEQ6L1 & AUET2B \cite{ATLASUETune1} \\
$\ttbar (Z/\gamma^*)$ + $\le 1$ parton & \textsc{MadGraph} & \textsc{Pythia} 6 & CTEQ6L1 & AUET2B \\
$t (Z/\gamma^*)$ & \textsc{MadGraph} & \textsc{Pythia} 6 & CTEQ6L1 & AUET2B \\
$q\bar q, qg \to WW, WZ$ & \textsc{Sherpa} \cite{sherpa} & \textsc{Sherpa} & CT10 & \textsc{Sherpa} default \\
$qq \to qqWW$, $qqWZ$, $qqZZ$ & \textsc{Sherpa} & \textsc{Sherpa} & CT10 & \textsc{Sherpa} default \\
$q\bar q, qg \to ZZ$ & \textsc{Powheg-BOX} \cite{powhegVV} & \textsc{Pythia} 8 & CT10 & AU2 \\
$gg \to ZZ$ & \textsc{gg2ZZ} \cite{Binoth:2008pr} & \textsc{Herwig} \cite{herwig} & CT10 & AUET2 \cite{ATLASUETune2} \\ 
$\ttbar$ & \textsc{Powheg-BOX} \cite{powhegtt} & \textsc{Pythia} 6 & CT10/CTEQ6L1 & Perugia2011C \cite{perugia} \\
$s$-, $t$-channel, $Wt$ single top & \textsc{Powheg-BOX} \cite{powhegstp,powhegstp2} & \textsc{Pythia} 6 & CT10/CTEQ6L1 & Perugia2011C \\
$Z \to \ell^+\ell^- + \le$ 5 partons & \textsc{Alpgen} \cite{alpgen} & \textsc{Pythia} 6 & CTEQ6L1 & Perugia2011C \\
$W \to \ell\nu + \le$ 5 partons & \textsc{Alpgen} & \textsc{Pythia} 6 & CTEQ6L1 & Perugia2011C \\
\hline\hline
 \end{tabular}
 }
\end{center}
\end{table*}

The event generator configurations used for simulating the signal and main background processes are shown in Table~\ref{tbl:evgen}.  Additional information is given below.

The \tth\ signal event simulation samples contain all Higgs boson decays with branching fractions set to values computed at NNLO in QCD \cite{Djouadi:1997yw,Bredenstein:2006rh,Actis:2008ts,Denner:2011mq,lhcxs}.  The  factorization ($\mu_{\rm F}$) and renormalization ($\mu_{\rm R}$) scales are set to
$m_t+\mH/2$. Higgs boson and top quark masses of 125 and 172.5 \gev, respectively, are used. 
These samples are the same as those used by other ATLAS \tth searches \cite{tthbb,2014tthdiph}.

Production of single top quarks with Higgs bosons is simulated as follows.  For $tHqb$, events are generated at leading order with \textsc{MadGraph} in the four-flavour scheme.  For $tHW$, events are generated at NLO with \textsc{MG5\_aMC@NLO} in the five-flavour scheme.  Higgs boson and top quark masses are set as for \tth production.

The main irreducible backgrounds are production of \ttw and \ttz (\ttv).  
For the \ttw\ process, events are generated at leading order with zero, one, or two extra partons in the final state, while for \ttz\
zero or one extra parton is generated. The important contribution from off-shell $\gamma^*/Z \to \ell^+\ell^-$ is 
included. The \tz\ process is simulated with the same setup, without extra partons.  

For diboson processes, the full matrix element for $\ell^+\ell^-$ production, including $\gamma^*$ and off-shell $Z$ contributions, is used.  The \textsc{Sherpa} $q\bar q$ and $qg$ samples include diagrams with additional partons in the final state at the matrix-element (ME) level, and include $b$- and $c$-quark mass effects.  \textsc{Sherpa} was found to have better agreement with data than \textsc{Powheg} for $WZ$, while the \textsc{Sherpa} and \textsc{Powheg} descriptions of $ZZ$ production are similar.

 A $t\bar{t}$+jets sample generated with the {\sc Powheg} NLO generator \cite{powhegtt} is used; the top quark mass is set to 172.5 \gev. Small corrections to the \ttbar system and top quark \pt spectra are applied based on discrepancies in differential distributions observed between data and simulation at 7 TeV~\cite{topdiff_7TEV}. Double-counting between the \ttbar\ and $Wt$ single top production final states is eliminated using the diagram-removal method~\cite{Frixione:2008yi}.

Samples of \zj and \wj events are generated with up to five additional partons using
the {\sc Alpgen v2.14}~\cite{alpgen} leading order (LO) generator. Samples are merged with matrix element-parton shower overlaps removed using MLM matching~\cite{mlm}.  Production of $b$- and $c$-quarks is also computed at matrix-element level, and overlaps between ME and parton shower production are handled by separating the kinematic regimes based on the angular separation of additional heavy partons.  The resulting ``light'' and ``heavy'' flavour samples are normalized by comparing the resulting $b$-tagged jet spectra with data.


All simulated samples with \textsc{Pythia}~6 and \textsc{Herwig} \cite{herwig} parton showering use {\sc Photos 2.15}~\cite{PhotosPaper} to model photon radiation and
{\sc Tauola 1.20}~\cite{TauolaPaper} for $\tau$ decays.  The \textsc{Herwig++} samples model photon radiation with \textsc{Photos} but use the internal $\tau$ decay model.  Samples using \textsc{Pythia}~8.1 and \textsc{Sherpa} use those generators' internal $\tau$ lepton decay and photon radiation generators. For \textsc{Herwig} samples, multiple parton interactions are modelled with \textsc{Jimmy} \cite{Butterworth:1996zw}.

Showered and hadronized events are passed through simulations of the ATLAS detector (either full {\sc GEANT4}~\cite{geant4} simulation or a hybrid simulation with parameterized calorimeter showers and GEANT4 simulation of the tracking systems~\cite{atlasSim,ATLAS:1300517}).
Additional minimum-bias $pp$ interactions (pileup) are modelled with the {\sc Pythia}~8.1 generator with the
{\sc MSTW2008} LO PDF set  and the A2 tune~\cite{ATLASUETune3}. They are added to the
signal and background simulated events according to the luminosity profile of the recorded data, with additional overall scaling to achieve a good match to observed calorimetry and tracking variables.
The contributions from pileup interactions both within the same bunch crossing as the hard-scattering process and in neighbouring bunch crossings are included in the simulation.

%% file: Objects.tex
\section{Object selection}

Electron candidates are reconstructed from
energy clusters in the electromagnetic calorimeter 
associated with reconstructed tracks in the inner detector. They are
required to have $|\eta_{\rm cluster}| < 2.47$. Candidates in the transition region $1.37 <
|\eta_{\rm cluster}| < 1.52$ between sections of the electromagnetic calorimeter are excluded. A multivariate discriminant based on shower shape and track information is used to distinguish electrons from hadronic showers~\cite{Aad:2014fxa,ATLAS-CONF-2014-032}.  Only electron candidates with transverse energy \et greater than 10 \gev\ are considered.
To reduce the background
from non-prompt electrons, i.e.~from decays of hadrons (including
heavy flavour) produced in jets, electron candidates are required
to be isolated. Two isolation variables, based on calorimetric and tracking variables, are computed.
The first ($E_\mathrm{T}^\mathrm{cone}$) is based on the sum of transverse energies of calorimeter cells within a cone of radius $\Delta R \equiv \sqrt{(\Delta\phi)^2 + (\Delta\eta)^2} = 0.2$ around the electron candidate direction.
This energy sum
excludes cells associated with the electron and is corrected
for leakage from the electromagnetic shower and ambient energy in the event. The second ($p_\mathrm{T}^\mathrm{cone}$) is defined 
based on tracks with $\pt > 1$~\gev\ within a cone 
of radius $\Delta R = 0.2$ around the electron candidate. Both isolation energies are separately required to be less than $0.05 \times  \et$. The longitudinal impact parameter of the electron track with respect
to the selected event primary vertex, multiplied by the sine of the polar angle, $|z_{0}\sin\theta|$, 
is required to be less
than 1~mm. The transverse impact parameter divided by the estimated uncertainty on its measurement, 
$|d_0|/\sigma(d_0)$, must be less than 4.  
If two electrons closer than $\Delta R=0.1$ are selected, only the one with the higher \pt\ is considered. 
An electron is rejected if, after passing all the above selections, it lies within $\Delta R=0.1$ 
of a selected muon.

Muon candidates are reconstructed by combining inner detector tracks with track segments or full tracks in the muon
spectrometer~\cite{mureco}. Only candidates with $|\eta|<2.5$ and $\pt > 10$~GeV are kept.
Additionally, muons are required to
be separated by at least $\Delta R > 0.04+(\mathrm{10}~\gev)/p_{\mathrm{T},\mu}$ from any selected jets (see below for details
on jet reconstruction and selection). The cut value is optimized to maximize the acceptance for prompt muons 
at a fixed rejection factor for non-prompt and fake muon candidates.
Furthermore, muons must satisfy similar $\ET^\mathrm{cone}$ and $\pt^\mathrm{cone}$ isolation criteria as for electrons, with both required to be less than $0.10 \times \pt$. 
The value of $|z_0\sin\theta|$ is required to be less than 1 mm, while $|d_0|/\sigma(d_0)$ must be less 
than 3.

Hadronically decaying $\tau$ candidates (\tauh) are reconstructed using clusters
in the electromagnetic and hadronic calorimeters.
The $\tau$ candidates are required to have \pt greater than 25~\gev\ and $|\eta|<2.47$.
The number of charged tracks associated with the $\tau$ candidates is
required to be one or three and
the charge of the $\tau$ candidates, determined from the associated
tracks, must be $\pm 1$.
The $\tau$ identification uses calorimeter cluster and tracking-based variables,
combined using
a boosted decision tree (BDT) \cite{ATLASTAUIDnew}.
An additional BDT which uses combined calorimeter and track quantities is employed 
to reject electrons reconstructed as one-prong hadronically decaying $\tau$ 
leptons.

Jets are reconstructed from calibrated topological
clusters~\cite{detectorpaper} built from energy deposits in the
calorimeters, using the anti-$k_t$
algorithm~\cite{ref:Cacciari2008,ref:Cacciari2006,ref:fastjet} with a
radius parameter $R=0.4$.  Prior to jet finding, a local cluster calibration
scheme~\cite{LCW1,LCW2} is applied to correct the topological cluster
energies for the effects of non-compensating calorimeter response, inactive material and
out-of-cluster leakage.
The jets are calibrated using energy and $\eta$-dependent calibration factors,
derived from simulations, to the mean energy of stable particles inside
the jets. Additional corrections to account for the difference between
simulation and data are derived from in-situ
techniques~\cite{JES, JER}.
After energy calibration, jets are required to have
\pt $>$ 25 \gev\ and $|\eta| <$ 2.5.

To reduce the contamination from jets originating in $pp$ interactions within the same
bunch crossing (pileup), the scalar sum of the $\pt$ of tracks matched to the jet 
and originating from the primary vertex
must be at least 50\% of the scalar sum of the $\pt$ of all tracks matched to the jet.
This criterion is only applied to jets with $\pt < 50 \gev$ (those most likely to originate from pileup) and $|\eta|<2.4$ (to avoid inefficiency at the edge of tracking acceptance).

The calorimeter energy deposits from electrons are typically also reconstructed as jets; in order to eliminate double counting, any jets within $\Delta R =$ 0.3
of a selected electron are not considered.

Jets containing $b$-hadrons are identified ($b$-tagged) via a multivariate discriminant~\cite{ATLAS-CONF-2014-046}
that combines information from the impact
parameters of displaced tracks with topological properties of
secondary and tertiary decay vertices reconstructed within the jet.
The working point used for this search corresponds to approximately 70\% efficiency to tag
a $b$-hadron jet, with a light-jet mistag rate of $\approx 1\%$ 
and a charm-jet rejection factor of 5,
as determined for $b$-tagged jets with \pt of 20--100 \gev\ and
$|\eta|<2.5$ in simulated $t\bar{t}$ events.  To avoid inefficiencies associated with the edge of the tracking coverage, only jets with $|\eta| < 2.4$ are considered as possible $b$-tagged jets in this analysis.  The efficiency and mistag rates of the $b$-tagging algorithm are measured in data \cite{ATLAS-CONF-2014-004,ATLAS-CONF-2014-046} and correction factors are applied to the simulated events.

%% file: Selection.tex
\section{Event selection and classification}

All events considered in this analysis are required to pass single-lepton ($e$ or $\mu$) triggers.  These achieve their maximal plateau efficiency for lepton $\pt > 25$ GeV.  

This analysis primarily targets the $H \to WW^*$ and $\tau\tau$ decay modes. Considering the decay of the \ttbar system as well, these \tth events contain either $WWWWb\bar b$ or $\tau\tau WWb\bar b$. The strategy is to target final states that cannot be produced in \ttbar decay alone --- i.e., three or more leptons, or two same-sign leptons --- thus suppressing  what would otherwise be the largest single background.

The analysis categories are classified by the number of light leptons and  hadronic $\tau$ decay candidates. The leptons are selected using the criteria described earlier. Events are initially classified by counting the number of light leptons with $\pt > 10\gev$. At least one light lepton is required to match a lepton selected by the trigger system. After initial sorting into analysis categories, in some cases the lepton selection criteria are tightened by raising the \pt threshold, tightening isolation selections or restricting the allowed $|\eta|$ range, as explained in the following per-category descriptions. The analysis includes five distinct categories: two same-sign light leptons with no \tauh (\twolnotau), three light leptons (\threel), two same-sign light leptons and one $\tauh$ (\twoltau), four light leptons (\fourl), and one light lepton and two $\tauh$ (\oneltwotau). The categories with \tauh candidates target the $H\to\tau\tau$ decay; the others are primarily sensitive to $H\to WW^*$ with a very small contribution from $H \to ZZ^*$.   The contributions to each category from different Higgs boson decay modes are shown in Table~\ref{tbl:contributions}. These selection criteria ensure that an event can only contribute to a single category. The contamination from gluon fusion, vector boson fusion, and associated $VH$ production mechanisms for the Higgs boson is predicted to be negligible.  Summed over all categories, the total expected number of reconstructed signal events assuming Standard Model \tth production is 10.2, corresponding to 0.40\% of all produced \tth events.  The detailed criteria for each category are described below. 

\begin{table}
\begin{center}
\caption{\label{tbl:contributions}Fraction of the expected \tth signal arising from different Higgs boson decay modes in each analysis category.  The six \twolnotau categories are combined together, as are the two \fourl categories. The decays contributing to the ``other'' column are dominantly $H\to\mu\mu$ and $H\to b\bar b$.  Rows may not add to 100\% due to rounding.}
 \begin{tabular}{lcccc}
  \hline\hline
 & \multicolumn{4}{c}{Higgs boson decay mode}\\
 Category & $WW^*$ & $\tau\tau$ & $ZZ^*$ & Other\\
  \hline
\twolnotau & 80\% & 15\% & \phantom{0}3\% & 2\%\\
\threel & 74\% & 15\% & \phantom{0}7\% & 4\% \\
\twoltau & 35\% & 62\% & \phantom{0}2\% & 1\% \\
\fourl & 69\% & 14\% & 14\% & 4\% \\
\oneltwotau & \phantom{0}4\% & 93\% & \phantom{0}0\% & 3\% \\
\hline\hline
 \end{tabular}
\end{center}
\end{table}

\subsection{$\twolnotau$ categories}
\label{subsec:2lepSS}

Selected events are required to include exactly two light leptons, which must have the same charge. Events with $\tauh$ candidates are vetoed.  To reduce the background from non-prompt leptons, the leading (subleading) lepton is required to satisfy $\pt > 25~(20)\gev$, and the muon isolation requirements are tightened to $E_{\text{T}}^{\text{cone}}/\pt<0.05$ and $\pt^{\text{cone}}/\pt<0.05$.  The angular acceptance of electron candidates is restricted to $|\eta| < 1.37$ in order to suppress $\ttbar$ background events where the sign of the electron charge is misreconstructed, as the charge misidentification rate increases at high pseudorapidity.

In order to suppress the lower-multiplicity \ttj and $\ttw$ backgrounds, events must include at least four reconstructed jets.  In order to suppress diboson and single-boson backgrounds, at least one of these jets must be $b$-tagged. The selected events are separated by lepton flavour ($e^{\pm}e^{\pm}$, $e^{\pm}\mu^{\pm}$, and $\mu^{\pm}\mu^{\pm}$) and number of jets (exactly four jets, at least five jets) into six categories with different signal-to-background ratio, resulting in higher overall sensitivity to the $\tth$ signal.
  
\subsection{$\threel$ category}
\label{subsec:3lep}

Selected events are required to include exactly three light leptons with total charge equal to $\pm 1$. Candidate events arising from non-prompt leptons overwhelmingly originate as opposite-sign dilepton events with one additional non-prompt lepton.  As a result, the non-prompt lepton is generally one of the two leptons with the same charge.  To reduce these backgrounds, a higher momentum threshold $\pt > 20$~GeV is applied to the two leptons with the same charge. No requirements are imposed on the number of $\tauh$ candidates. In order to suppress the \ttj and $\ttv$ backgrounds,  selected events are required to include either at least four jets of which at least one must be $b$-tagged, or exactly three jets of which at least two are $b$-tagged. To suppress the $\ttz$ background, events that contain an opposite-sign same-flavour lepton pair with the dilepton invariant mass within 10~GeV of the $Z$ mass are vetoed.  Events containing an opposite-sign lepton pair with invariant mass below $12\gev$ are also removed to suppress background from resonances that decay to light leptons.

\subsection{$\twoltau$ category}
\label{subsec:2lepSStau}

Selected events are required to include exactly two light leptons, with the same charge and leading (subleading) $\pt >$ 25~(15) \gev, and exactly one hadronic $\tau$ candidate. The reconstructed charge of the $\tauh$ candidate has to be opposite to that of the light leptons. In order to reduce \ttj and $\ttv$ backgrounds, events must include at least four reconstructed jets.  In order to suppress diboson and single-boson backgrounds, at least one jet must be $b$-tagged. To suppress the \zj background, events with dielectron invariant mass within 10~GeV of the $Z$ mass are vetoed. 

\subsection{$\fourl$ categories}
\label{subsec:4lep}

Selected events are required to include exactly four light leptons with total charge equal to zero and leading (subleading) $\pt > 25~(15)~$~GeV. No requirements are applied on the number of $\tauh$ candidates. In order to suppress the \ttj and $\ttv$ backgrounds, the selected events are required to include at least two jets of which at least one must be $b$-tagged. To suppress the $\ttz$ background, events that contain an opposite-sign same-flavour lepton pair with dilepton invariant mass within 10~GeV of the $Z$ mass are vetoed. In order to suppress background contributions from resonances that decay to light leptons, all opposite-sign same-flavour lepton pairs are required to have a dilepton invariant mass greater than 10~GeV. The four-lepton invariant mass is required to be between 100 and 500~GeV, which gives high acceptance for \tth, $H \to WW^* \to \ell\nu\ell\nu$, but rejects $Z \to 4\ell$ and high-mass \ttz events. Selected events are separated by the presence or absence of a same-flavour, opposite-sign lepton pair into two categories, referred to respectively as the $Z$-enriched and $Z$-depleted categories. In both cases the $Z$ mass veto is applied, but background events in the $Z$-enriched category can arise from off-shell $Z$ and $\gamma^* \to \ell^+\ell^-$ processes while in the $Z$-depleted category these backgrounds are absent.

\subsection{$\oneltwotau$ category} 
\label{subsec:1lep2tau} 

Selected events are required to include exactly one light lepton with $\pt > 25\gev$ and exactly two hadronic $\tau$ candidates. The $\tauh$ candidates must have opposite charge. In order to suppress the \ttj and $\ttv$ backgrounds, events must include at least three reconstructed jets. In order to suppress diboson and single-boson backgrounds, at least one of the jets must be $b$-tagged.  This final state is primarily sensitive to $H \to \tau^+\tau^-$ decays, allowing use of the invariant mass of the visible decay products of the $\tauh\tauh$ system ($m_\mathrm{vis}$) as a signal discriminant.  Signal events are required to satisfy $60 < m_\mathrm{vis} < 120\gev$.

%% file: Background.tex
\section{Background estimation} \label{sec:bkgest}

Important irreducible backgrounds include $\ttv$ and diboson production and are estimated from MC simulation.  Validation regions enriched in these backgrounds are used to verify proper modelling of data by simulation.  Reducible backgrounds are due to non-prompt lepton production and electron charge mis-identification, and are estimated from data, with input from simulation in some categories.  In the \oneltwotau category the primary concern is fake \tauh candidates, which are modelled using simulation and validated against a data-driven estimate.

\begin{figure*}
 \begin{center}
\subfloat{\includegraphics[width=0.5\linewidth]{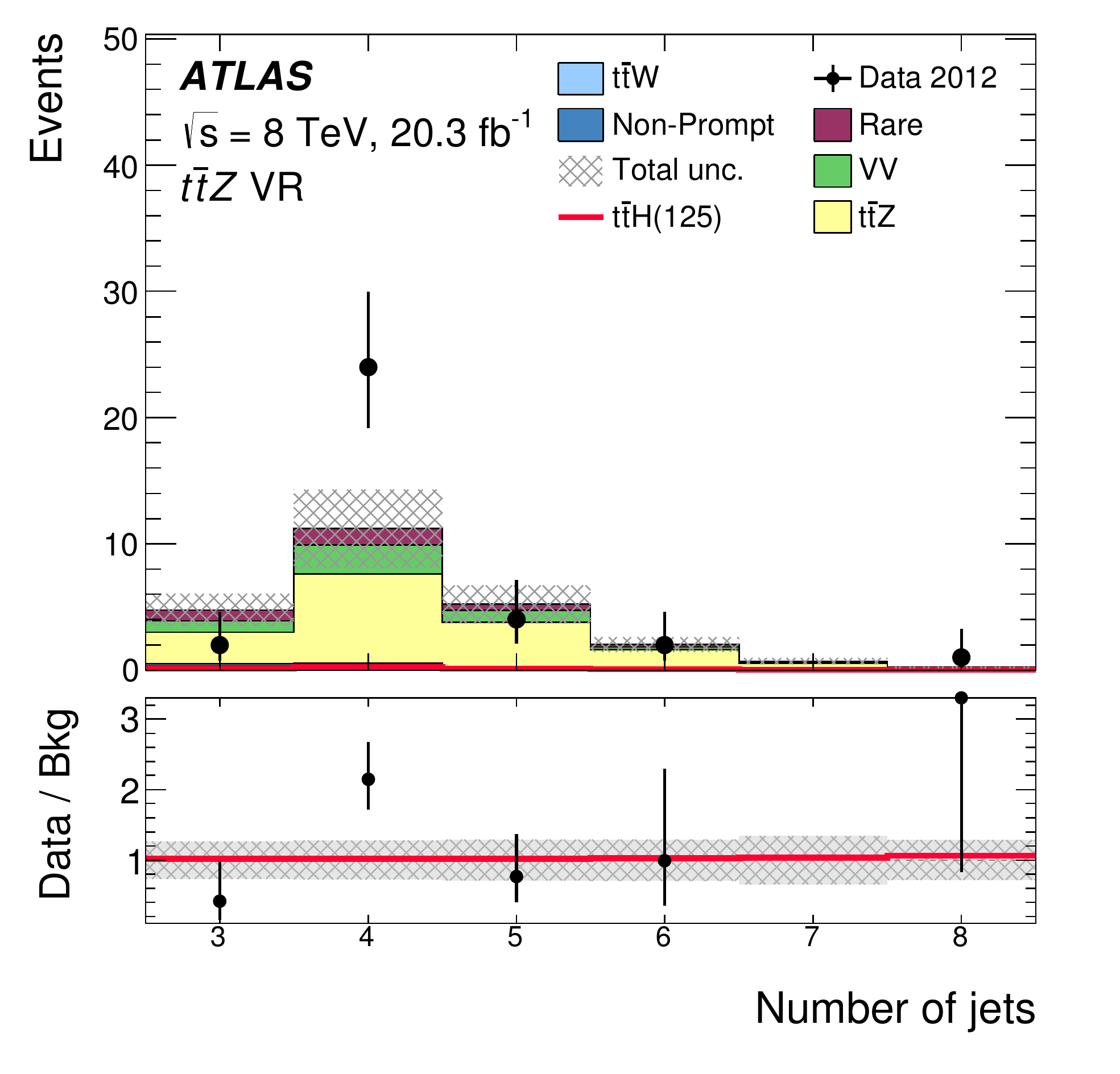}}%
\subfloat{\includegraphics[width=0.5\linewidth]{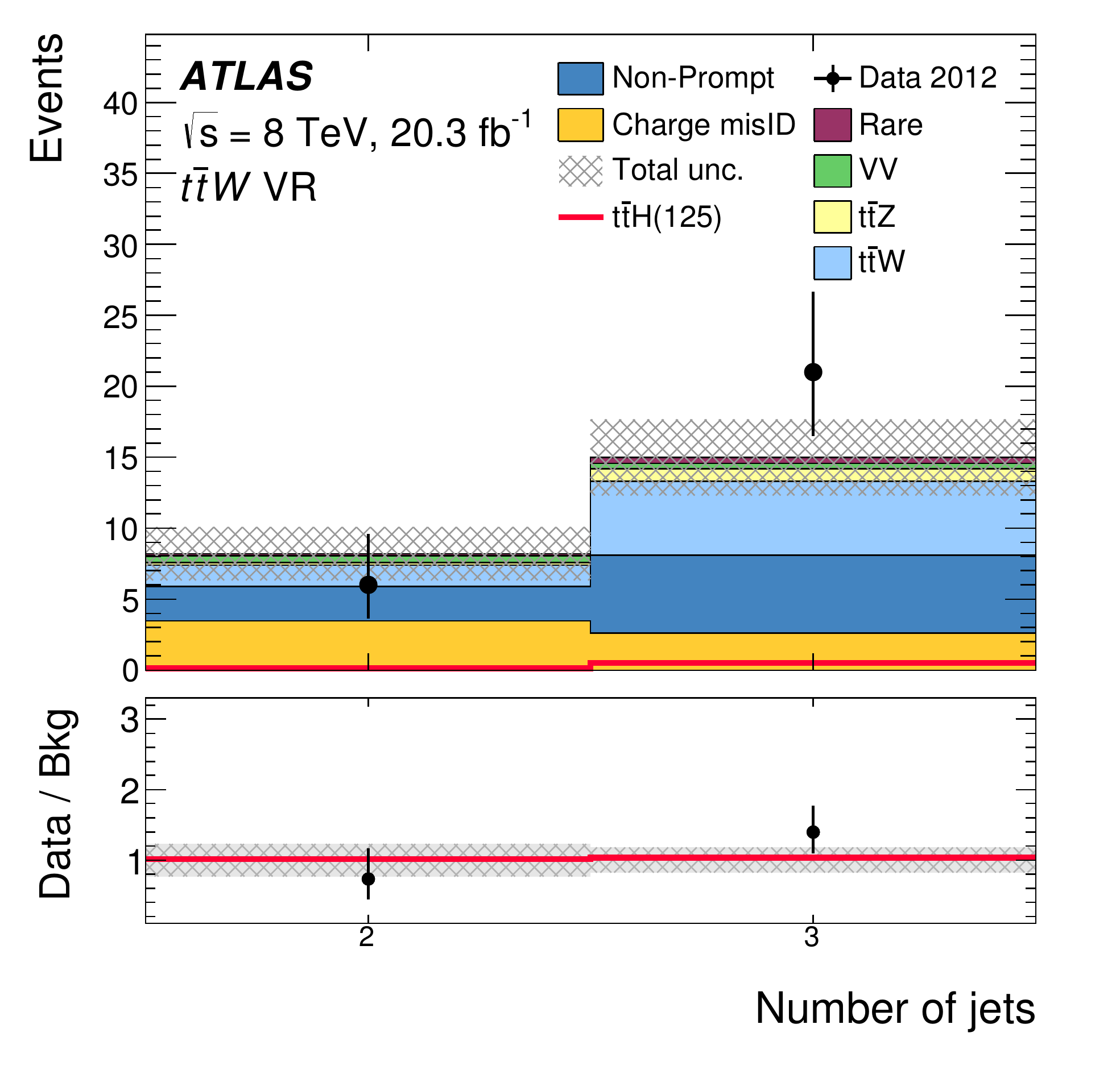}}
\caption{\label{fig:vrplots}The spectrum of the number of jets expected and observed in the \ttz (left) and \ttw (right) validation regions (VR).  The hatched band represents the total uncertainty on the background prediction in each bin.  The ``non-prompt'' backgrounds are those with a lepton arising from a hadron decay or from a photon conversion in detector material.  Rare processes include $t Z$, $t\bar t WW$, triboson, $t\bar t t\bar t$, and $t H$ production. The overlaid red line corresponds to the \tth signal predicted by the SM.}
 \end{center}
\end{figure*}

\subsection{\ttv and \tz}
The primary backgrounds with prompt leptons stem from the production of \ttw and \ttz.  The \ttw background tends to have lower jet multiplicity than the signal and so the leading contribution comes from events with additional high-\pt jets; it is the major \ttv contribution in the \twolnotau categories and comparable to \ttz in the \twoltau category.  The \ttz process has similar multiplicity to the \tth signal but can only contribute to the signal categories when the \Zboson~boson decays leptonically, so the on-shell contribution can be removed by vetoing events with opposite-sign dilepton pairs with invariant mass near the \Zboson pole.  This is the larger of the two \ttv contributions for the \threel, \fourl, and \oneltwotau categories.  The \tz process makes a subleading contribution to both channels.  A validation region is used to verify the modelling of \ttz using on-shell \Zboson decays.  Agreement is seen within the large statistical uncertainty.  No region of equivalent purity and statistical power exists for \ttw production; nevertheless the expectations are cross-checked with a validation region defined with the \twolnotau selection except with two or more $b$-tagged jets and either two or three jets, where the \ttw purity is $\approx 30\%$, and are found to be consistent within uncertainties.  The spectra of the number of jets in these validation regions is shown in Fig.~\ref{fig:vrplots}. 

Uncertainties on the \ttv background contributions arise from both the overall cross section uncertainties (see Section~\ref{ssec:xsections}) and the acceptance uncertainties.  The latter are estimated by comparing particle-level samples after showering produced by three different pairs of generators: a) the nominal \textsc{MadGraph} LO merged sample versus an equivalent LO merged sample generated with \textsc{Sherpa} 2.1.1, to account for ME-parton shower matching effects; b) the LO merged \textsc{Sherpa} sample versus a \textsc{Sherpa+OpenLoops} \cite{openloops} NLO sample, to compare LO merged and NLO acceptance; and c) \textsc{MG5\_aMC@NLO} with \textsc{Pythia} 8 parton shower versus \textsc{Herwig++} parton shower, to compare \pt-ordered versus angular-ordered parton showers.  Each of these variations is input independently into the final fit.  When summed in quadrature they have an impact of 5--23\% depending on the category and background source (\ttw versus \ttz).  Uncertainties arising from changes in the acceptance due to the choice of QCD scale and PDF are also evaluated; these have an impact of 1.3--6.7\% for scale and 0.9--4.8\% for PDF.

\subsection{\label{sec:otherpromptbkg}Other prompt lepton contributions}
Other backgrounds with prompt leptons arise from multiboson processes ($WZ$, $ZZ$, and triboson production) in association with heavy-flavour jets, or with a misidentified light-flavour jet.  The main process affecting the final result is $WZ$ + jets.  Validation regions with three leptons including a \Zboson candidate and either zero or one $b$-tagged jet are studied. The number of jets in $WZ+0b$ events is reproduced well in the highly populated bins (up to 4 jets), leading to the conclusion that the jet radiation spectrum is well modelled.  The dominant uncertainty on the prediction in the signal region is expected to arise from the $WZ+b$ cross section.  Data constrain this component with roughly 100\% uncertainty.    As a result a 100\% uncertainty is assigned to the $WZ+b$ cross section, giving a 50\% uncertainty on the total $WZ$ yield, correlated across categories.  The cross sections for production of $WW+b$ and $ZZ+b$ are also assigned 50\% uncertainties; these have negligible impact on the final result.

\subsection{Charge sign misidentification}
The process $e^\pm \to e^\pm \gamma \to e^\pm e^+ e^-$ occurring in detector material can result in an electron produced with nearly the same momentum as the parent electron but with opposite charge. In these cases the observed electron has opposite charge to that of the primary electron (charge mis-id). The analogous processes $\mu^\pm \to \mu^\pm e^+ e^-$ and $\mu^{\pm} \rightarrow \mu^{\pm} \mu^{+} \mu^{-}$ have negligible rates for the selected events. The \ttbar{} and \zdyj events that undergo this process contribute to \twolnotau in the $ee$ and $e\mu$ categories. As electrons pass through more material at high $|\eta|$, the charge mis-id rate increases as well, and so the electron $|\eta| < 1.37$ requirement significantly reduces the impact of this background.  The charge mis-id rate due to track curvature mismeasurement for electrons and muons is negligible. 

The charge mis-id probability is determined by a maximum-likelihood fit using $Z \to ee$ events reconstructed as same-sign and as opposite-sign pairs, as a function of electron $\eta$ and \pt. This probability function is then applied to a sample of events passing the \twolnotau selection except that the lepton pair is required to be opposite sign. The charge mis-id probability from the relatively low momentum $Z$ daughters is extrapolated to higher \pt using scaling functions extracted from Monte Carlo simulations.  The dominant uncertainty is due to the statistical precision of the charge mis-id probability determination, and is $\approx 40\%$ in the signal regions.

\subsection{Non-prompt light leptons}
A significant background arises from leptons not produced in decays of electroweak bosons (non-prompt leptons), which can promote (for example) a single-lepton \ttbar event into a \twolnotau category or a dilepton \ttbar event to the \threel or \twoltau categories.  These backgrounds in the signal regions are expected to be dominated by \ttbar{} or single top quark production with leptons produced in decays of heavy-flavour hadrons. Production of \ttbar with an additional photon which converts in the detector material is a subdominant contribution. With the tight object selection requirements applied in this analysis, almost all reconstructed electron and muon objects correspond to real electrons and muons; the fraction arising from incorrect particle identification is negligible. 
Estimates of these backgrounds are obtained from data.  Each channel has a slightly different procedure, motivated by the specific event topology and the statistical power available in the control regions.  The methods are discussed below, and the expected non-prompt lepton contributions to the various categories are shown in Table~\ref{tbl:yields}.  In the following, a \textit{tight} lepton is a lepton that passes the nominal selection, a \textit{sideband} lepton is defined as a lepton candidate which satisfies different criteria  than the tight lepton selection (identification selection, isolation, or \pt), and (sideband) \textit{control regions} either require one or more sideband leptons to replace a tight lepton in the signal region selection, or have the same lepton selection as the signal region but different jet requirements.

\input{YieldTable}

\subsubsection{\twolnotau categories}
The non-prompt lepton yields in the signal regions are estimated by extrapolating from sideband control regions in data which are enriched in \ttbar non-prompt contributions.  For electrons, sideband objects are selected by inverting the electron identification and isolation requirements; for muons the sideband objects have low transverse momentum, $6 < \pt < 10\gev$, but otherwise are selected the same way as nominal muons.  Transfer factors are used to extrapolate from events with one tight and one sideband lepton, but which otherwise pass the signal region selections, to the signal regions with two tight leptons.  These transfer factors are determined from additional data control regions (tight + sideband and two tight leptons) with lower jet multiplicity ($1 \le n_\mathrm{jet} \le 3$ for electrons, $2 \le n_\mathrm{jet} \le 3$ for muons).  In all regions the expected contribution from processes producing prompt leptons is subtracted before extracting transfer factors or using the yields for extrapolation.   For channels with electrons, the charge mis-id background is also subtracted, and a dilepton mass veto is applied in the control regions to suppress contributions from $Z\to e^+ e^-$ decays.  A cross-check on the muon estimate, using an extrapolation in muon isolation instead of muon \pt, agrees well with the nominal procedure and provides additional confidence in the estimate.

The systematic uncertainties on this procedure are estimated by checking a) its ability to successfully predict the non-prompt background in \ttbar simulation and b) the stability of the prediction using data when the selection of the control regions is altered.  For the former, different parton shower and $b$-hadron decay models were checked, as was the result of removing the $b$-tagged jet requirement. In addition, for electrons, the effects of relaxing the pseudorapidity requirement to $|\eta| < 2.5$ and of raising the \pt threshold were studied.  These checks show stability at the 25--30\% level, limited by the statistical precision of the simulations.  
The stability in data is checked by altering the \pt required for the $b$-tagged jet, applying a requirement on missing transverse momentum\footnote{This is calculated using calorimeter energy deposits, calibrated according to associated reconstructed physics objects, and also including the transverse momenta of reconstructed muons.} \met, extracting the transfer factors only from events with three jets, or (for muons) using 10--15~\gev\ muons as the sideband objects.   This check shows stability of the predictions to 14\% for muons and 19\% for electrons.    Additional systematic uncertainties in the prediction arise from the statistical uncertainties on the yields in the control regions and the subtraction of prompt and charge mis-id contributions.  The overall uncertainties on the non-prompt yield prediction in any given category range from 32\% to 52\%, and correlations between the categories due to uncertainties in the transfer factors are included in the fit (see Section~\ref{sec:results}). 

\subsubsection{\threel category}
Sideband leptons are defined by reversing the isolation requirement for electrons and muons and, for electrons, requiring that the candidate fail the tight electron identification discriminant requirement of the analysis but pass a looser selection.  The non-prompt lepton contribution in the signal region is estimated by extrapolating from data regions with two tight and one sideband lepton, using transfer factors estimated from Monte Carlo simulation.   These events typically contain two prompt opposite-sign leptons and one non-prompt lepton, which necessarily must be of the same sign as one of the prompt leptons.  Therefore the non-prompt lepton estimation procedure is applied only to the two same-sign leptons.  The simulation-derived transfer factor is validated in a region of lower jet multiplicity ($2 \le \njet \le 3$ and exactly one $b$-tagged jet). Good agreement is observed in this validation region between the prediction (11.8 $\pm$ 2.3) and the observed yield (9.8 $\pm$ 4.9 events after prompt background subtraction).  Systematic uncertainties in the procedure are derived by studying the agreement between data and simulation in the variables used for the extrapolation, which is $\approx 20$\% for both electrons and muons.  Additional uncertainties arise from the statistical uncertainties on the yields in the control regions and in the \ttbar simulation. 

\subsubsection{\label{sssec:faketwoltau}\twoltau category}
Reconstructing two same-sign light leptons from \ttbar production or similar sources requires that one of the light leptons is non-prompt or has its charge misidentified.  In the \twoltau category, the charge mis-id contribution is negligible and the primary concern is non-prompt light leptons.  Around half of the \tauh candidates in these events come from $W\to \tau \nu$ decays, while the remainder arise from misidentified light-quark or gluon jets.  Regardless of whether the \tauh candidate is a fake, there is also a non-prompt light lepton.  Due to this fact, sidebands in the light-lepton selection criteria are used, analogously to the \twolnotau and \threel categories.   Since the ratio of real and fake \tauh candidates is similar in the signal and all control regions, fake \tauh candidates are not accounted for separately; the small variations in the ratio in the control regions are found to have negligible impact on the total estimate in the signal region.  In order to maintain similar origin composition of the non-prompt leptons, the \et isolation requirement is inverted, the \pt isolation requirement is relaxed, and for electrons the identification criteria are also relaxed to a looser working point.  The low jet multiplicity region $2 \le \njet \le 3$ is used to determine a transfer factor from sideband to tight lepton selections.  The expected non-prompt lepton yield in the signal region is obtained by using this transfer factor to extrapolate from a control region with the same jet selection as the signal region but with one tight and one sideband light lepton.    The procedure is validated by checking that it correctly reproduces the signal region yield expected in \ttbar simulations.  The assigned systematic uncertainty (27\%) is dominated by the statistical precision of this test.  The overall uncertainty on the non-prompt background prediction is dominated by the limited statistics of the high jet multiplicity control region.

\subsubsection{\fourl category}
The non-prompt lepton contribution in this category is expected to be negligible and is estimated to be $\lesssim 10^{-3}$ events in the $Z$-enriched sample and $\lesssim 10^{-4}$ events in the $Z$-depleted sample.  In both cases this represents $\lesssim 2\%$ of the total background expectation. These estimates are obtained using the transfer factors from the \threel channel and appropriate control regions with two loose leptons and relaxed jet multiplicity requirements.

\subsection{\tauh misidentification in the \oneltwotau category}
The nominal estimate for the fake \tauh yield is derived from \ttbar simulation.  To obtain a sufficiently large sample size, fast simulation using parameterized calorimeter showers is used.   At all preselection stages the simulation is found to give an acceptable description of the \ttbar background, both in kinematic distributions and total yield.  This estimate is cross-checked with the data-driven method described below.  

Of the two \tauh candidates, one is opposite in sign to the light lepton (OS) and the other has the same sign (SS).  The SS candidate is almost always a fake \tauh, while the light lepton is prompt and the OS \tauh candidate is often real ($\approx 30\%$). A sideband \tauh is defined as a candidate passing a loose identification BDT selection but not the nominal tight one. Assuming the \tauh candidate fake probabilities are not correlated between jets identified as OS and SS candidates, control regions can be used to predict yields in the signal region.  There are three control regions, depending on whether only the OS, only the SS, or both the OS and SS \tauh candidates are sideband objects.  The two regions with sideband OS \tauh candidates are used to obtain the transfer factor for the SS \tauh candidate, which is then applied to the region with a tight OS and sideband SS candidate to obtain the prediction for the signal region where both are tight. The transfer factor is measured as a function of the \pt, $\eta$, number of tracks, and $b$-tag discriminant value of the SS \tauh candidate.  The data-driven method is cross-checked in \ttbar simulation and found to successfully predict the yields in the signal region.  The main limitation of this method is the statistical power of the control regions.

The simulation-driven method is taken as the primary estimate, as the validation of the method at preselection stages is more precise than the data-driven method due to larger event yield for the former.  The comparison of the simulation- and data-driven techniques gives a 36\% uncertainty in the prediction in the signal region, which is taken as the systematic uncertainty on the estimate.

%% file: YieldTable.tex
\begin{table*}
\begin{center}
\caption{\label{tbl:yields} Expected and observed yields in each channel. Uncertainties shown are the sum in quadrature of systematic uncertainties and Monte Carlo simulation statistical uncertainties.  ``Non-prompt'' includes the misidentified \tauh background to the \oneltwotau category.  Rare processes ($tZ$, $\ttbar WW$, triboson production, $t\bar t t \bar t$, $tH$) are not shown as a separate column but are included in the total expected background estimate.}
{\small
 \begin{tabular}{lccccccccccc}
\hline\hline
Category & $q$ mis-id & Non-prompt & \ttw & \ttz & Diboson & Expected bkg. & \tth $(\mu=1)$ & Observed\\
\hline
  $ee$ + $\ge 5j$  & 1.1 $\pm$ 0.5 & 2.3 $\pm$ 1.2 & 1.4 $\pm$ 0.4 & 0.98 $\pm$ 0.26 & 0.47 $\pm$ 0.29 & 6.5 $\pm$ 1.8 & 0.73 $\pm$ 0.14 & \obsjust{10}\\
 $e\mu$ + $\ge 5j$  & 0.85 $\pm$ 0.35 & 6.7 $\pm$ 2.4 & 4.8 $\pm$ 1.2 & 2.1 $\pm$ 0.5 & 0.38 $\pm$ 0.30 & 15 $\pm$ 3\phantom{0} & 2.13 $\pm$ 0.41 & \obsjust{22}\\
 $\mu\mu$ + $\ge 5j$ & -- & 2.9 $\pm$ 1.4 & 3.8 $\pm$ 0.9 & 0.95 $\pm$ 0.25 & 0.69 $\pm$ 0.39 & 8.6 $\pm$ 2.2 & 1.41 $\pm$ 0.28 &\obsjust{11}\\
 $ee$ + $4j$ & 1.8 $\pm$ 0.7 & 3.4 $\pm$ 1.7 & 2.0 $\pm$ 0.4 & 0.75 $\pm$ 0.20 & 0.74 $\pm$ 0.42 & 9.1 $\pm$ 2.1 & 0.44 $\pm$ 0.06 &  \obsjust{9}\\
 $e\mu$ + $4j$ & 1.4 $\pm$ 0.6 & 12 $\pm$ 4\phantom{0} & 6.2 $\pm$ 1.0 & 1.5 $\pm$ 0.3 & 1.9 $\pm$ 1.0 & 24 $\pm$ 5\phantom{0} & 1.16 $\pm$ 0.14 & \obsjust{26}\\
 $\mu\mu$ + $4j$ & -- & 6.3 $\pm$ 2.6 & 4.7 $\pm$ 0.9 & 0.80 $\pm$ 0.22 & 0.53 $\pm$ 0.30  & 12.7 $\pm$ 2.9\phantom{0} & 0.74 $\pm$ 0.10 & \obsjust{20}\\
 \threel & -- & 3.2 $\pm$ 0.7 & 2.3 $\pm$ 0.7 & 3.9 $\pm$ 0.8 & 0.86 $\pm$ 0.55 & 11.4 $\pm$ 2.3\phantom{0} & 2.34 $\pm$ 0.35 & \obsjust{18}\\ 
 \twoltau & -- & 0.4 $^{+0.6}_{-0.4}$  & 0.38 $\pm$ 0.12 & 0.37 $\pm$ 0.08 & 0.12 $\pm$ 0.11 & 1.4 $\pm$ 0.6 & 0.47 $\pm$ 0.08 & \obsjust{1} \\ 
 \oneltwotau & -- & 15 $\pm$ 5\phantom{0} & 0.17 $\pm$ 0.06 & 0.37 $\pm$ 0.09 & 0.41 $\pm$ 0.42 & 16 $\pm$ 5\phantom{0} & 0.68 $\pm$ 0.13 & \obsjust{10}\\
 \fourl $Z$-enr. & -- & $\lesssim 10^{-3}$ & $\lesssim 3 \times 10^{-3}$ & 0.43 $\pm$ 0.12 & 0.05 $\pm$ 0.02 & 0.55 $\pm$ 0.15 & 0.17 $\pm$ 0.02 & \obsjust{1}\\
 \fourl $Z$-dep. & -- & $\lesssim 10^{-4}$ & $\lesssim 10^{-3}$ & 0.002 $\pm$ 0.002 & $\lesssim 2\times 10^{-5}$ & 0.007 $\pm$ 0.005 & 0.025 $\pm$ 0.003 & \obsjust{0}\\
\hline\hline
 \end{tabular}
 }
\end{center}
\end{table*}

%% file: Syst.tex
\section{Other systematic uncertainties}
Systematic uncertainties not already discussed are summarized below.

 The uncertainty on the integrated luminosity is 2.8\%.  This uncertainty is derived from a calibration of the luminosity scale derived from beam-separation scans performed in November 2012, following the same methodology as that detailed in Ref.~\cite{Aad:2013ucp}.

Lepton reconstruction and identification uncertainties are obtained from $Z\to \ell\ell$, $\Zboson\to \ell\ell\gamma$, $\Upsilon \to \ell\ell$, and $J/\psi \to \ell\ell$ events \cite{Aad:2014fxa,ATLAS-CONF-2014-032,mureco}. Uncertainties on the detector response are assessed similarly to other ATLAS analyses.  The modelling of the efficiency of the tight isolation requirements in simulation is explicitly checked as a function of the number of jets in the event.  These corrections are found to be very small, with uncertainties limited by data statistics.

The largest jet-related systematic uncertainty arises from the jet energy scale, in particular contributions from the in-situ calibration in data, the different response to quark and gluon jets, and the pileup subtraction.  The impact of the $b$-tagging efficiency uncertainty on the signal strength $\mu = \sigma_{\tth, \mathrm{obs}}/\sigma_{\tth, \mathrm{SM}}$ at the best-fit value of $\mu$ is $\Delta \mu = ^{+0.08}_{-0.06}$. Because only one (of typically two) $b$-jets present in signal or \ttv events is required to be tagged, the uncertainty on the $b$-tagging efficiency (while included) does not have as large an effect in this analysis as it does in other \tth searches such as those targeting the $H\to b\bar b$ decay.  
 
The uncertainties on the inclusive \tth production cross section are discussed in Section~\ref{ssec:xsections}.  Additionally, the effects of PDF uncertainty, QCD scale choice, and parton shower algorithm on the signal acceptance in each analysis category are considered.  The resulting relative uncertainties on the acceptance are 0.3--1.4\% for PDF, 0.1--2.7\% for scale choice, and 1.5--13\% for parton shower algorithm.

For most backgrounds the uncertainties from Monte Carlo simulation sample sizes are negligible.  For the diboson backgrounds, however, these can reach 50\% of the total diboson yield uncertainties shown in Table~\ref{tbl:yields}.

%% file: Results.tex
\section{\label{sec:results}Results}
\input{MajorFloats}
The observed yields, and a comparison with the expected backgrounds and \tth signal, are shown in Table~\ref{tbl:yields}.  The distributions of the number of jets in the events passing signal region selections are shown in Fig.~\ref{fig:jetplots}.  The best-fit value of the signal strength $\mu = \sigma_{\tth, \mathrm{obs}}/\sigma_{\tth, \mathrm{SM}}$ is determined using a maximum likelihood fit to the data yields of the categories listed in Table~\ref{tbl:yields}, which are treated as independent Poisson terms in the likelihood.  The fit is based on the profile-likelihood approach where the systematic uncertainties are treated as nuisance parameters with prior uncertainties that can be further constrained by the fit \cite{asym}. The $\mu=1$ hypothesis assumes Standard Model Higgs boson production and decay with $m_H = 125\gev$; for all other values of $\mu$ only the \tth production cross section is scaled (the Higgs boson branching fractions are fixed to their SM values).  

Systematic uncertainties are allowed to float in the fit as nuisance parameters and take on their best-fit values.  The only constraints on nuisance parameter uncertainties found by the fit are for non-prompt lepton transfer factors and normalization region yields in the \twolnotau categories and the fake \tauh background yield in the \oneltwotau category.  The former all have large statistical components and so the additional information from the signal regions is expected to constrain them.  The latter has a very large initial uncertainty which the fit is able to constrain as $\mu$ is required to be the same in all categories.  The largest difference between pre- and post-fit nuisance parameter values is in the \oneltwotau fake estimate, which shifts by $-1.0\sigma$ due to the deficit of observed relative to expected events.  The next largest effect is a $+0.4\sigma$ shift in the \twolnotau non-prompt $\mu$ transfer factor.

\begin{table}
 \begin{center}
 \caption{\label{tbl:systematics} Leading sources of systematic uncertainty and their impact on the measured value of $\mu$.}
  \begin{tabular}{lcc}
   \hline\hline
   Source & \multicolumn{2}{c}{$\Delta \mu$}\\
   \hline
   \twolnotau non-prompt muon transfer factor & $+0.38$ & $-0.35$\\
   \ttw acceptance & ${+0.26}$ & ${-0.21}$ \\
   \tth inclusive cross section & ${+0.28}$  &${-0.15}$ \\
   Jet energy scale & ${+0.24}$ & ${-0.18}$\\
   \twolnotau non-prompt electron transfer factor & ${+0.26}$ & ${-0.16}$\\
   \tth acceptance & ${+0.22}$ & ${-0.15}$\\
   \ttz inclusive cross section & ${+0.19}$ & ${-0.17}$\\
   \ttw inclusive cross section & ${+0.18}$ & ${-0.15}$\\
   Muon isolation efficiency & ${+0.19}$ & ${-0.14}$\\
   Luminosity & ${+0.18}$ & ${-0.14}$\\
   \hline\hline
  \end{tabular}

 \end{center}

\end{table}

\renewcommand{\obsjust}[1]{\makebox[0.7cm]{\null\hfill{}#1}}
\begin{table*}
 \begin{center}
 \caption{\label{tbl:limits} Observed and expected 95\% CL upper limits, derived using the CL$_s$ method, on the strength parameter $\mu = \sigma_{\tth, \mathrm{obs}}/\sigma_{\tth, \mathrm{SM}}$ for a Higgs boson of mass $m_H = 125$~\gev. The last column shows the median expected limit in the presence of a \tth signal of Standard Model strength.}
  \begin{tabular}{c|c|ccccc|c}
  \hline\hline
  & & \multicolumn{6}{c}{Expected Limit}\\
  Channel & Observed Limit & $-2\sigma$ & $-1\sigma$ & Median & $+1\sigma$ & $+2\sigma$ & Median ($\mu=1$)\\
  \hline
   \twolnotau & \obsjust{\phantom{0}6.7} & \obsjust{2.1} & \obsjust{2.8} & \obsjust{3.9} & \obsjust{5.7} & \obsjust{8.4} & \obsjust{5.0} \\
   \threel & \obsjust{6.8} & \obsjust{2.0} & \obsjust{2.7} & \obsjust{3.8} & \obsjust{5.7} & \obsjust{8.5} & \obsjust{5.1} \\
   \twoltau & \obsjust{7.5} & \obsjust{4.5} & \obsjust{6.1} & \obsjust{8.4} & \obsjust{13\phantom{.0}} & \obsjust{21\phantom{.0}} & \obsjust{10\phantom{.0}} \\
   \fourl & \obsjust{18\phantom{.0}} & \obsjust{\phantom{0}8.0} & \obsjust{11\phantom{.0}} & \obsjust{15\phantom{.0}} & \obsjust{23\phantom{.0}} & \obsjust{39\phantom{.0}} & \obsjust{17\phantom{.0}} \\
   \oneltwotau & \obsjust{13\phantom{.0}} & \obsjust{10\phantom{.0}} & 13\phantom{.0} & 18\phantom{.0} & 26\phantom{.0} & 40\phantom{.0} & \obsjust{19\phantom{.0}} \\
   \hline
   Combined & \obsjust{4.7} & \obsjust{1.3} & \obsjust{1.8} & \obsjust{2.4} & \obsjust{3.6} & \obsjust{5.3} & \obsjust{3.7} \\
   \hline\hline
  \end{tabular}
 \end{center}
\end{table*}

\begin{figure}
\begin{center}
 \includegraphics[width=\linewidth]{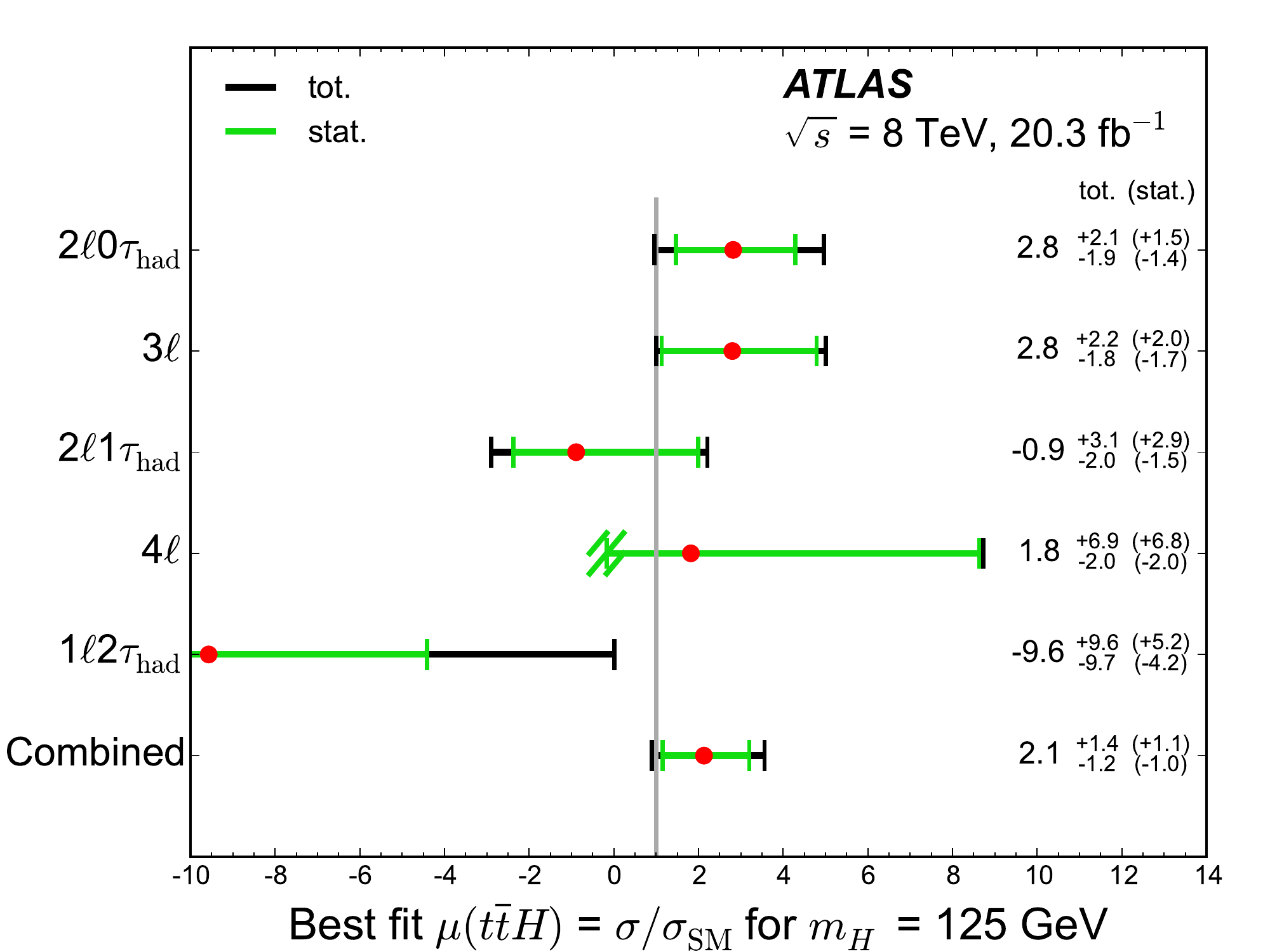}
 \caption{\label{fig:fits} Best-fit values of the signal strength parameter $\mu = \sigma_{\tth, \mathrm{obs}}/\sigma_{\tth, \mathrm{SM}}$.  For the $4\ell$ $Z$-depleted category, $\mu < -0.17$ results in a negative expected total yield and so the lower uncertainty is truncated at this point. }
 \end{center}
\end{figure}

The results of the fit are shown in Fig.~\ref{fig:fits}.  The impact of the most important systematic uncertainties on the measured value of $\mu$ in the combined fit is shown in Table~\ref{tbl:systematics}.  In each category, the uncertainties on $\mu$ are mainly statistical, except for the combined \twolnotau result where the statistical and systematic uncertainties are of comparable size.  In the \fourl $Z$-depleted category, a (non-physical) signal strength $\mu < -0.17$ results in a negative expected total yield and a discontinuity in the profiled likelihood; the error bar is therefore truncated at this point. The results are compatible with the Standard Model expectation and with previous searches for \tth production in multilepton final states \cite{Khachatryan:2014qaa}.  Combined over all categories, the value of $\mu$ is found to be $2.1 ^{+1.4}_{-1.2}$.  In the presence of a signal of SM strength, the combined fit is expected to return $\mu = 1.0 ^{+1.2}_{-1.1}$. The $\mu=0$ hypothesis has an observed (expected) $p$-value of 0.037 (0.18), corresponding to $1.8\sigma$ ($0.9\sigma$).  The $\mu=1$ hypothesis (the SM) has an observed $p$-value of 0.18, corresponding to $0.9\sigma$.  The likelihood function can be used to obtain 95\% confidence level (CL) upper limits on $\mu$ using the CL$_s$ method \cite{cls,asym}, leading to the results in Table~\ref{tbl:limits}.  The observed (expected) upper limit, combining all channels, is $\mu <$ 4.7 (2.4).

This analysis is a search for \tth production; as such, production of $tHqb$ and $tHW$ is considered as a background and set to Standard Model expectation.  Including this contribution as a background induces a shift of $\Delta \mu = -0.04$ compared to setting it to zero.  
A full extraction of limits on the top quark Yukawa coupling including the relevant modifications of single top plus Higgs boson production is reported in Ref.~\cite{ATLAScouplings}.

The results are sensitive to the assumed cross sections for \ttw and \ttz production, and use theoretical predictions for these values as experimental measurements do not yet have sufficient precision.  The best-fit $\mu$ value as a function of these cross sections is

\[ \mu(\tth) = 2.1 -1.4 \left(\frac{\sigma(\ttw)}{232\textrm{ fb}}-1\right) -1.3 \left(\frac{\sigma(\ttz)}{206\textrm{ fb}}-1\right) \]

%% file: MajorFloats.tex
\begin{figure*}
 \begin{center}
\subfloat[Combined \twolnotau categories]{\includegraphics[width=.5\linewidth]{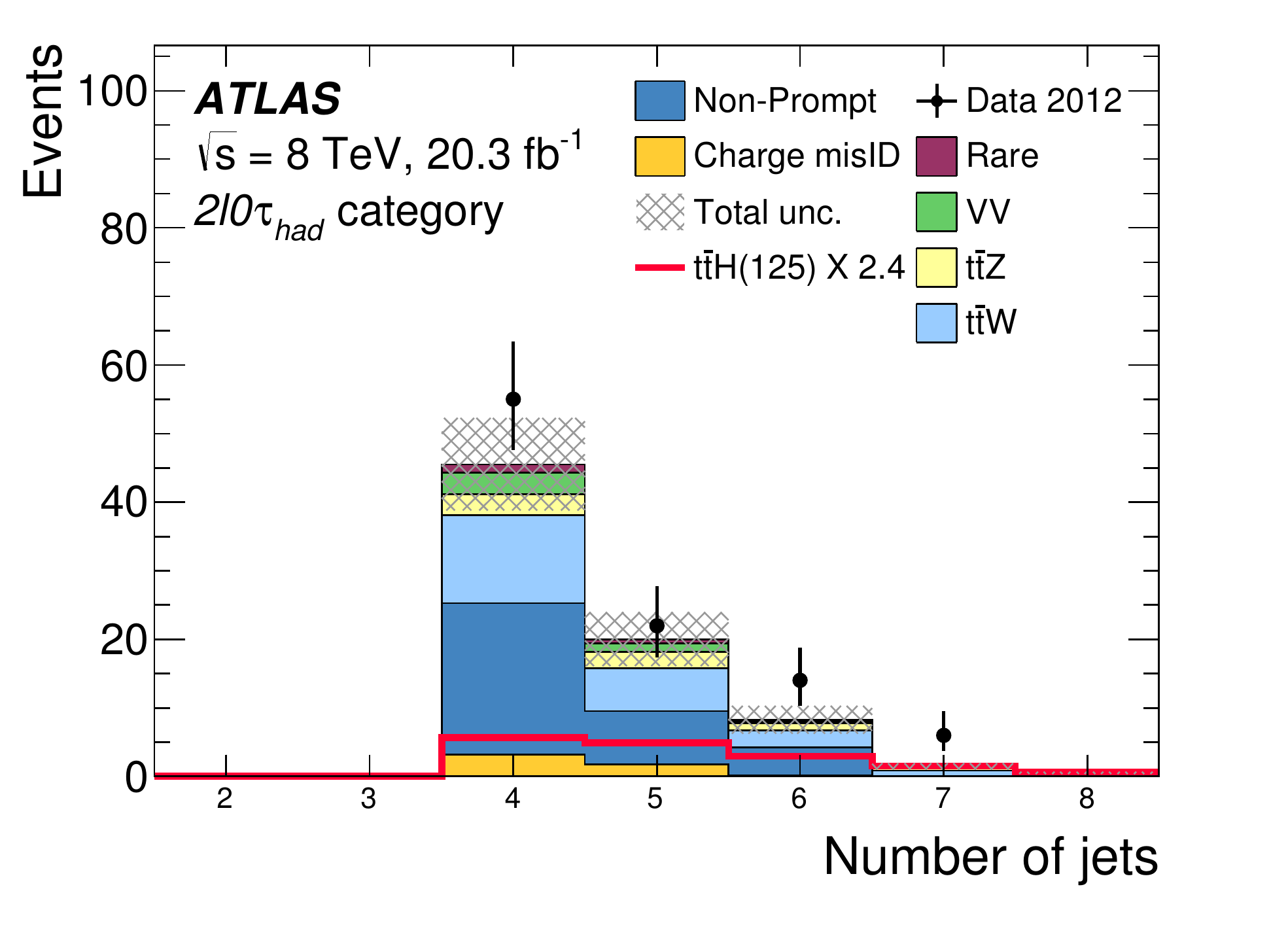}}%
\subfloat[\threel category]{\includegraphics[width=.5\linewidth]{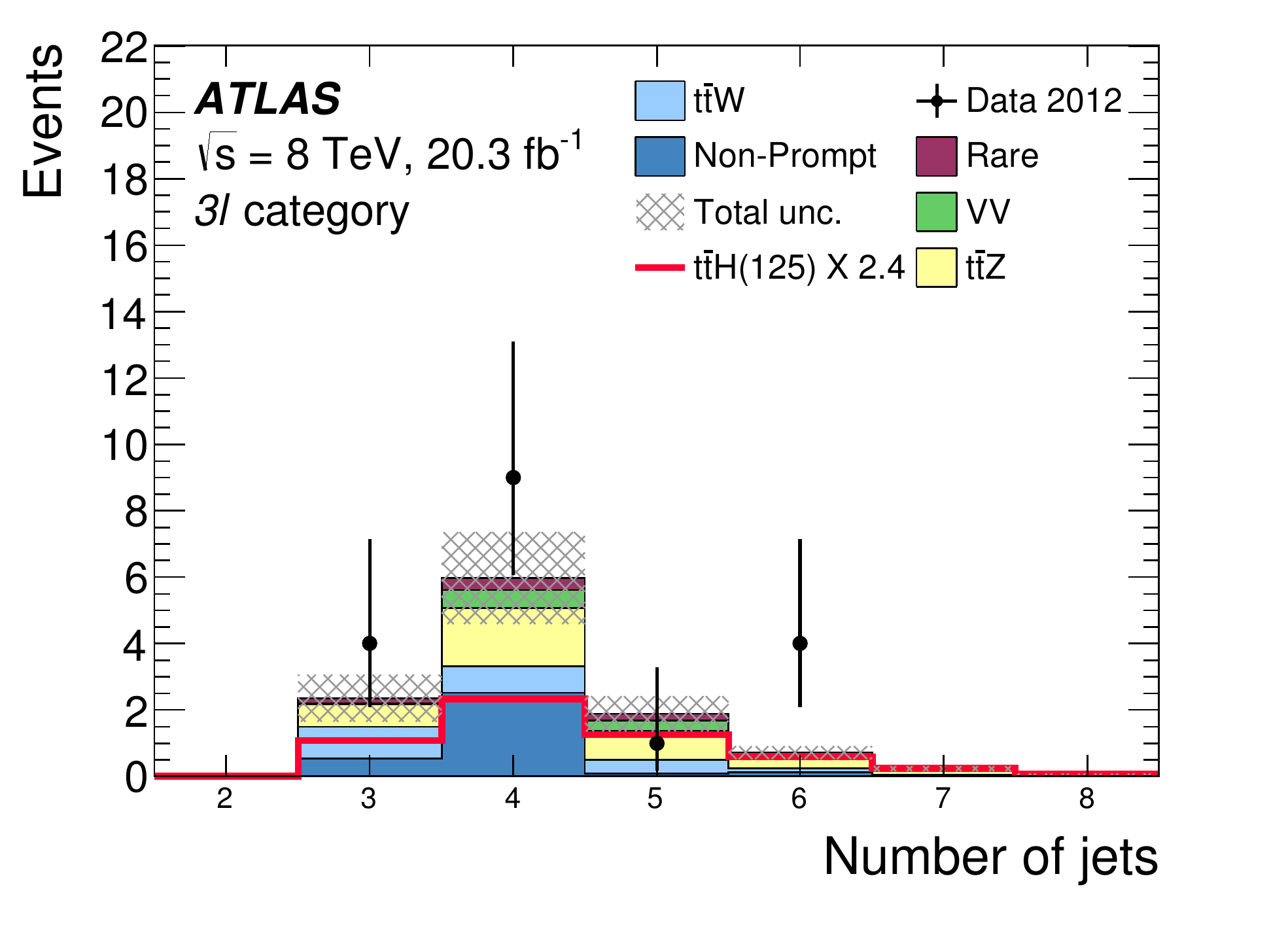}}\\
\subfloat[\twoltau category]{\includegraphics[width=.5\linewidth]{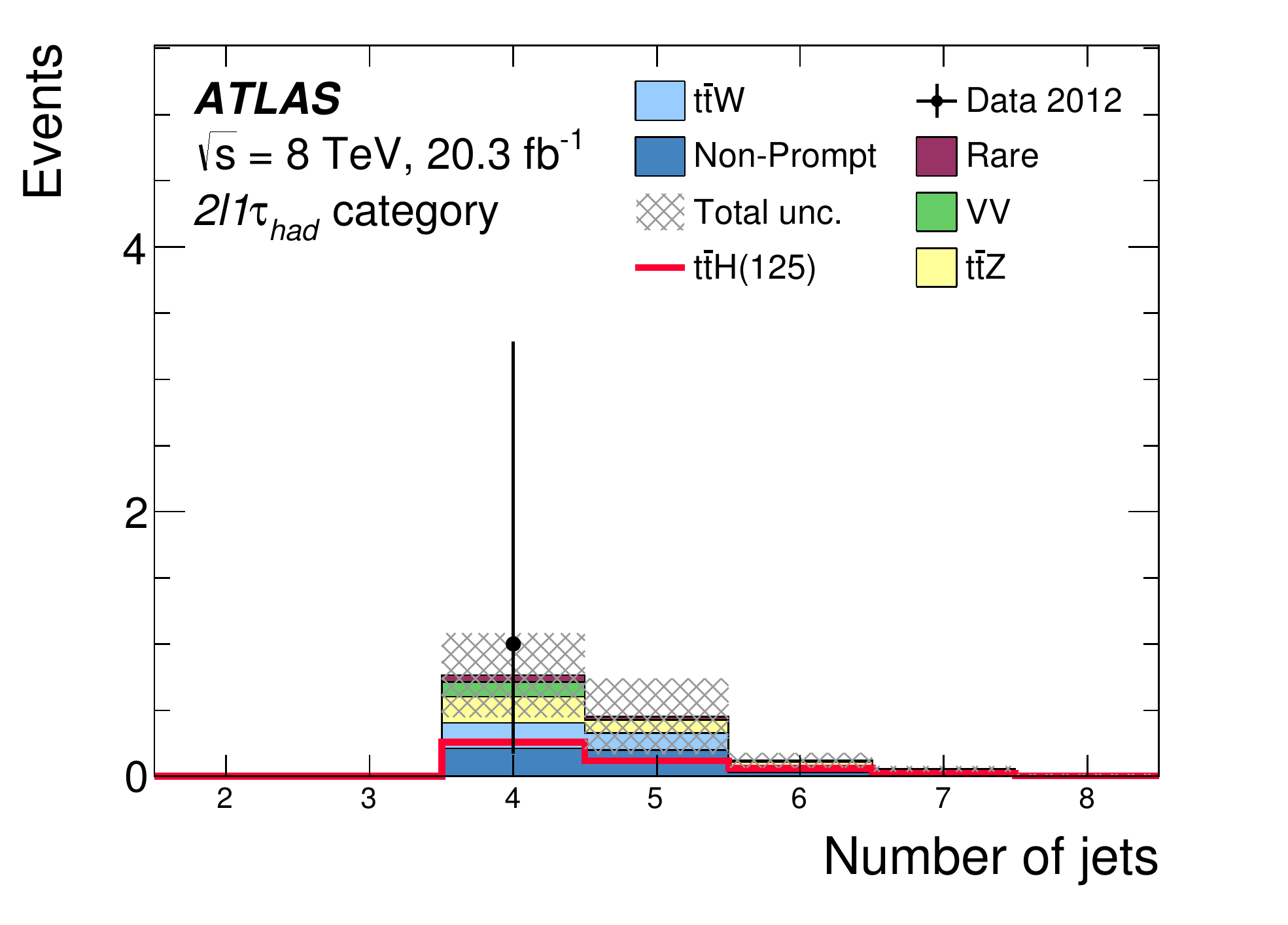}}%
\subfloat[Combined \fourl categories]{\includegraphics[width=.5\linewidth]{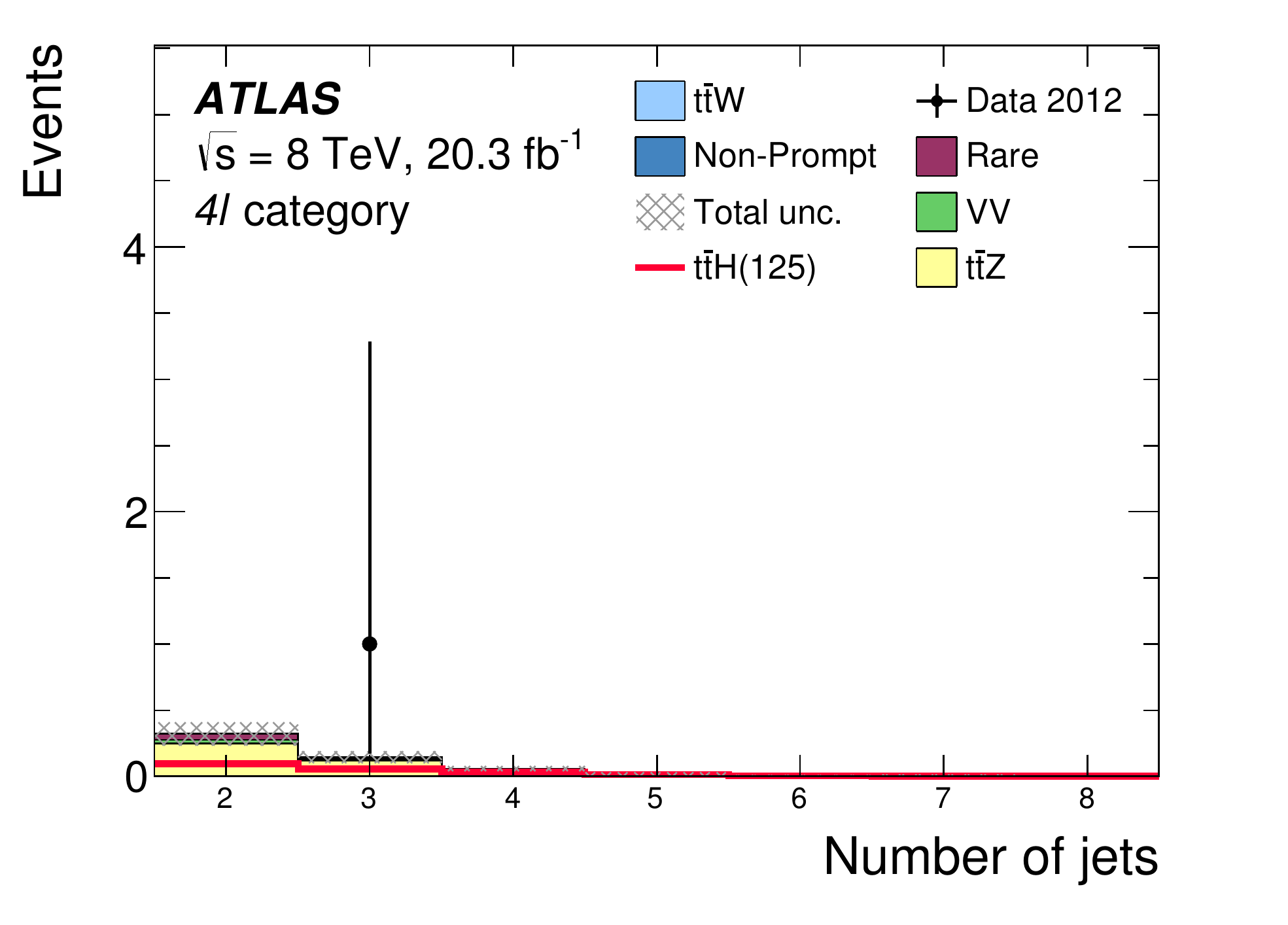}}\\
\subfloat[\oneltwotau category]{\includegraphics[width=.5\linewidth]{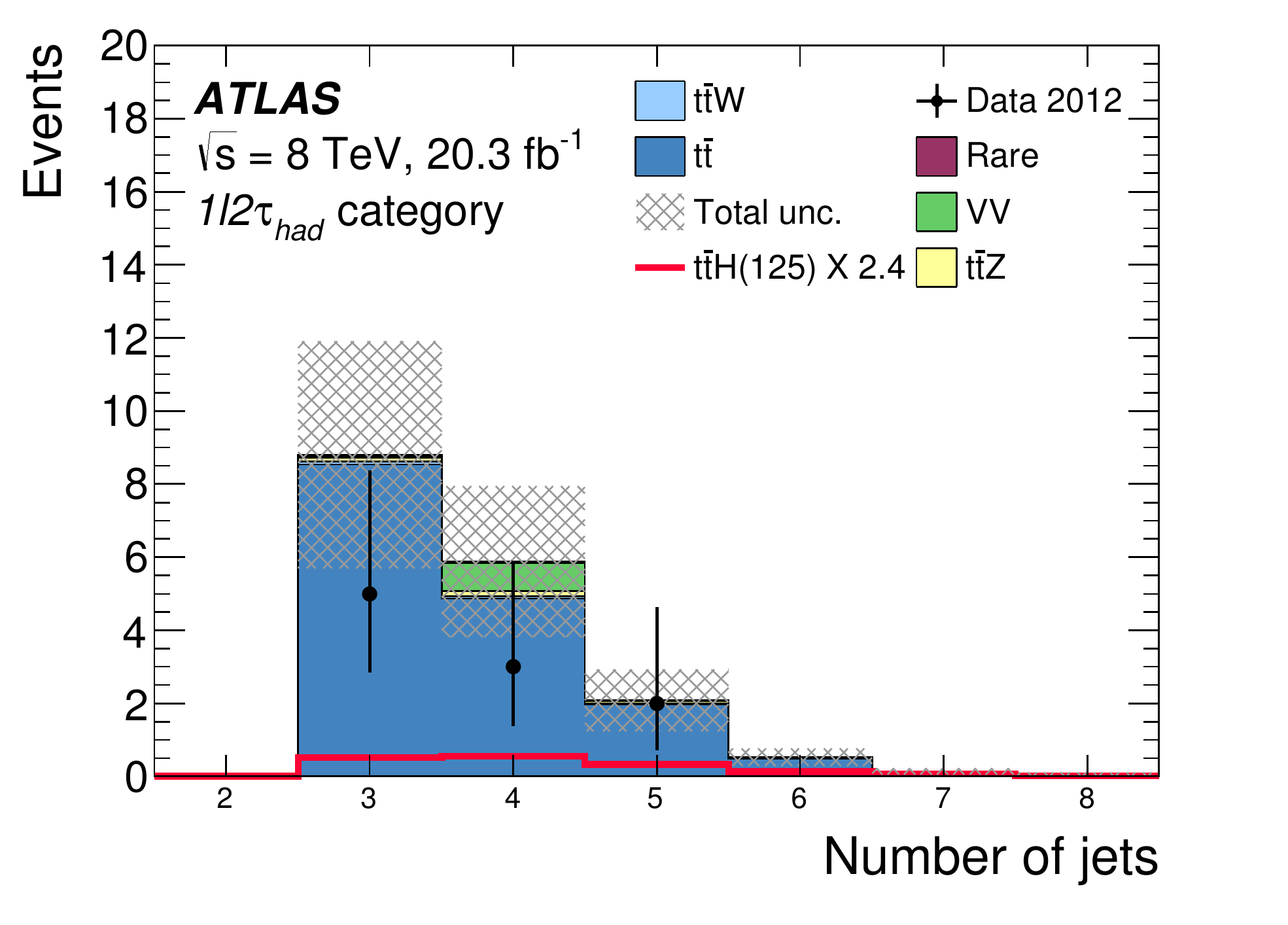}}
\caption{\label{fig:jetplots}The spectrum of the number of jets expected and observed in each signal region.  For display purposes the six \twolnotau categories ($ee$/$e\mu$/$\mu\mu$ and $=4$/$\ge 5$ jets) are combined into one plot, as are the two \fourl categories ($Z$-enriched and $Z$-depleted). The hatched bands show the total uncertainty on the background prediction in each bin.  The non-prompt and charge mis-id background spectra are taken from simulation of \ttbar, single top, \zj, and other small backgrounds, with normalization as described in the text (in particular the $=4$/$\ge 5$ jet regions of the \twolnotau plot have the ratio given by the data-driven prediction). The overlaid red line shows the \tth signal from the Standard Model.  For visibility, the \tth signal is multiplied by a factor of 2.4 in the \twolnotau, \threel, and \oneltwotau plots.}
 \end{center}

\end{figure*}

%% file: Conclusions.tex
\section{Conclusions}

A search for \tth production in multilepton final states has been performed using 20.3~fb$^{-1}$ of proton--proton collision data at $\sqrt{s}=8$~TeV recorded by the ATLAS experiment at the LHC.  The best-fit value of the ratio $\mu$ of the observed production rate to that predicted by the Standard Model is $2.1 ^{+1.4}_{-1.2}$.    This result is consistent with the Standard Model expectation. A 95\% confidence level limit of $\mu < 4.7$ is set.  The expected limit in the absence of \tth signal is $\mu < 2.4$. The observed (expected) $p$-value of the no-signal hypothesis corresponds to $1.8\sigma$ ($0.9\sigma$).

%% file: Acknowledgements.tex

We thank CERN for the very successful operation of the LHC, as well as the
support staff from our institutions without whom ATLAS could not be
operated efficiently.

We acknowledge the support of ANPCyT, Argentina; YerPhI, Armenia; ARC,
Australia; BMWFW and FWF, Austria; ANAS, Azerbaijan; SSTC, Belarus; CNPq and FAPESP,
Brazil; NSERC, NRC and CFI, Canada; CERN; CONICYT, Chile; CAS, MOST and NSFC,
China; COLCIENCIAS, Colombia; MSMT CR, MPO CR and VSC CR, Czech Republic;
DNRF, DNSRC and Lundbeck Foundation, Denmark; EPLANET, ERC and NSRF, European Union;
IN2P3-CNRS, CEA-DSM/IRFU, France; GNSF, Georgia; BMBF, DFG, HGF, MPG and AvH
Foundation, Germany; GSRT and NSRF, Greece; RGC, Hong Kong SAR, China; ISF, MINERVA, GIF, I-CORE and Benoziyo Center, Israel; INFN, Italy; MEXT and JSPS, Japan; CNRST, Morocco; FOM and NWO, Netherlands; BRF and RCN, Norway; MNiSW and NCN, Poland; GRICES and FCT, Portugal; MNE/IFA, Romania; MES of Russia and NRC KI, Russian Federation; JINR; MSTD,
Serbia; MSSR, Slovakia; ARRS and MIZ\v{S}, Slovenia; DST/NRF, South Africa;
MINECO, Spain; SRC and Wallenberg Foundation, Sweden; SER, SNSF and Cantons of
Bern and Geneva, Switzerland; NSC, Taiwan; TAEK, Turkey; STFC, the Royal
Society and Leverhulme Trust, United Kingdom; DOE and NSF, United States of
America.

The crucial computing support from all WLCG partners is acknowledged
gratefully, in particular from CERN and the ATLAS Tier-1 facilities at
TRIUMF (Canada), NDGF (Denmark, Norway, Sweden), CC-IN2P3 (France),
KIT/GridKA (Germany), INFN-CNAF (Italy), NL-T1 (Netherlands), PIC (Spain),
ASGC (Taiwan), RAL (UK) and BNL (USA) and in the Tier-2 facilities
worldwide.

%% file: atlas_authlist.tex
\begin{flushleft}
{\Large The ATLAS Collaboration}

\bigskip

G.~Aad$^{\rm 85}$,
B.~Abbott$^{\rm 113}$,
J.~Abdallah$^{\rm 151}$,
O.~Abdinov$^{\rm 11}$,
R.~Aben$^{\rm 107}$,
M.~Abolins$^{\rm 90}$,
O.S.~AbouZeid$^{\rm 158}$,
H.~Abramowicz$^{\rm 153}$,
H.~Abreu$^{\rm 152}$,
R.~Abreu$^{\rm 30}$
% ...
\bigskip
\\
% ...
$^{11}$ Institute of Physics, Azerbaijan Academy of Sciences, Baku, Azerbaijan\\
% ...
$^{30}$ CERN, Geneva, Switzerland\\
% ...
$^{85}$ CPPM, Aix-Marseille Universit{\'e} and CNRS/IN2P3, Marseille, France\\
% ...
$^{90}$ Department of Physics and Astronomy, Michigan State University, East Lansing MI, United States of America\\
% ...
$^{107}$ Nikhef National Institute for Subatomic Physics and University of Amsterdam, Amsterdam, Netherlands\\
% ...
$^{113}$ Homer L. Dodge Department of Physics and Astronomy, University of Oklahoma, Norman OK, United States of America\\
% ...
$^{151}$ Institute of Physics, Academia Sinica, Taipei, Taiwan\\
$^{152}$ Department of Physics, Technion: Israel Institute of Technology, Haifa, Israel\\
$^{153}$ Raymond and Beverly Sackler School of Physics and Astronomy, Tel Aviv University, Tel Aviv, Israel\\
% ...
$^{158}$ Department of Physics, University of Toronto, Toronto ON, Canada\\
% ...
\end{flushleft}


%% file: ttHmultilep.bbl
\providecommand{\href}[2]{#2}\begingroup\raggedright\begin{thebibliography}{10}

\bibitem{Glashow:1961tr}
S.~L. Glashow, {\em Partial Symmetries of Weak Interactions},
\href{http://dx.doi.org/10.1016/0029-5582(61)90469-2}{Nucl. Phys. {\bfseries
  22} (1961) 579}.

\bibitem{Weinberg:1967tq}
S.~Weinberg, {\em A Model of Leptons},
\href{http://dx.doi.org/10.1103/PhysRevLett.19.1264}{Phys. Rev. Lett.
  {\bfseries 19} (1967) 1264}.

\bibitem{sm_salam}
A.~Salam, {\em ``Weak and electromagnetic interactions,'' in Elementary
  particle theory: relativistic groups and analyticity}.
\newblock N. Svartholm, ed., p. 367. Alm\-qvist \& Wiksell, Stockholm, 1968.
\newblock Proceedings of the eighth Nobel symposium.

\bibitem{Englert:1964et}
F.~Englert and R.~Brout, {\em Broken symmetry and the mass of gauge vector
  mesons},
\href{http://dx.doi.org/10.1103/PhysRevLett.13.321}{Phys. Rev. Lett. {\bfseries
  13} (1964) 321}.

\bibitem{higgs12}
P.~W. Higgs, {\em Broken symmetries, massless particles and gauge fields},
\href{http://dx.doi.org/10.1016/0031-9163(64)91136-9}{Phys. Rev. Lett.
  {\bfseries 12} (1964) 132}.

\bibitem{Higgs:1964pj}
P.~W. Higgs, {\em Broken symmetries and the masses of gauge bosons},
\href{http://dx.doi.org/10.1103/PhysRevLett.13.508}{Phys. Rev. Lett. {\bfseries
  13} (1964) 508}.

\bibitem{ghk}
G.~Guralnik, C.~R. Hagen, and T.~W.~B. Kibble, {\em Global conservation laws
  and massless particles},
  \href{http://dx.doi.org/10.1103/PhysRevLett.13.585}{Phys.~Rev.~Lett.
  {\bfseries 13} (1964) 585}.

\bibitem{2012gk}
{ATLAS} Collaboration, {\em Observation of a new particle in the search for the
  Standard Model Higgs boson with the ATLAS detector at the LHC},
  \href{http://dx.doi.org/10.1016/j.physletb.2012.08.020}{Phys. Lett. B
  {\bfseries 716} (2012) 1},
\href{http://arxiv.org/abs/1207.7214}{{\ttfamily arXiv:1207.7214 [hep-ex]}}.

\bibitem{2012gu}
{CMS} Collaboration, {\em Observation of a new boson at a mass of 125 GeV with
  the CMS experiment at the LHC},
  \href{http://dx.doi.org/10.1016/j.physletb.2012.08.021}{Phys. Lett. B
  {\bfseries 716} (2012) 30},
\href{http://arxiv.org/abs/1207.7235}{{\ttfamily arXiv:1207.7235 [hep-ex]}}.

\bibitem{Aad:2014eha}
{ATLAS} Collaboration, {\em {Measurement of Higgs boson production in the
  diphoton decay channel in pp collisions at center-of-mass energies of 7 and 8
  TeV with the ATLAS detector}},
  \href{http://dx.doi.org/10.1103/PhysRevD.90.112015}{Phys.~Rev.~D {\bfseries
  90} (2014) 112015},
\href{http://arxiv.org/abs/1408.7084}{{\ttfamily arXiv:1408.7084 [hep-ex]}}.

\bibitem{Khachatryan:2014ira}
{CMS} Collaboration, {\em {Observation of the diphoton decay of the Higgs boson
  and measurement of its properties}},
  \href{http://dx.doi.org/10.1140/epjc/s10052-014-3076-z}{Eur.~Phys.~J.~C
  {\bfseries 74} (2014) 3076},
\href{http://arxiv.org/abs/1407.0558}{{\ttfamily arXiv:1407.0558 [hep-ex]}}.

\bibitem{Aad:2014eva}
{ATLAS} Collaboration, {\em {Measurements of Higgs boson production and
  couplings in the four-lepton channel in pp collisions at center-of-mass
  energies of 7 and 8 TeV with the ATLAS detector}},
  \href{http://dx.doi.org/10.1103/PhysRevD.91.012006}{Phys.~Rev.~D {\bfseries
  91} (2015) 012006},
\href{http://arxiv.org/abs/1408.5191}{{\ttfamily arXiv:1408.5191 [hep-ex]}}.

\bibitem{Chatrchyan:2013mxa}
{CMS} Collaboration, {\em {Measurement of the properties of a Higgs boson in
  the four-lepton final state}},
  \href{http://dx.doi.org/10.1103/PhysRevD.89.092007}{Phys.~Rev.~D {\bfseries
  89} (2014) 092007},
\href{http://arxiv.org/abs/1312.5353}{{\ttfamily arXiv:1312.5353 [hep-ex]}}.

\bibitem{ATLAS:2014aga}
{ATLAS} Collaboration, {\em {Observation and measurement of Higgs boson decays
  to WW$^*$ with the ATLAS detector}},
  \href{http://dx.doi.org/10.1103/PhysRevD.92.012006}{Phys.~Rev.~D {\bfseries
  92} (2015) 012006},
\href{http://arxiv.org/abs/1412.2641}{{\ttfamily arXiv:1412.2641 [hep-ex]}}.

\bibitem{Chatrchyan:2013iaa}
{CMS} Collaboration, {\em {Measurement of Higgs boson production and properties
  in the WW decay channel with leptonic final states}},
  \href{http://dx.doi.org/10.1007/JHEP01(2014)096}{JHEP {\bfseries 1401} (2014)
  096},
\href{http://arxiv.org/abs/1312.1129}{{\ttfamily arXiv:1312.1129 [hep-ex]}}.

\bibitem{Aad:2015vsa}
{ATLAS} Collaboration, {\em {Evidence for the Higgs-boson Yukawa coupling to
  tau leptons with the ATLAS detector}},
  \href{http://dx.doi.org/10.1007/JHEP04(2015)117}{JHEP {\bfseries 1504} (2015)
  117},
\href{http://arxiv.org/abs/1501.04943}{{\ttfamily arXiv:1501.04943 [hep-ex]}}.

\bibitem{Chatrchyan:2014nva}
{CMS} Collaboration, {\em {Evidence for the 125 GeV Higgs boson decaying to a
  pair of $\tau$ leptons}},
  \href{http://dx.doi.org/10.1007/JHEP05(2014)104}{JHEP {\bfseries 1405} (2014)
  104},
\href{http://arxiv.org/abs/1401.5041}{{\ttfamily arXiv:1401.5041 [hep-ex]}}.

\bibitem{Khachatryan:2014qaa}
{CMS} Collaboration, {\em Search for the associated production of the Higgs
  boson with a top-quark pair},
  \href{http://dx.doi.org/10.1007/JHEP09(2014)087}{JHEP {\bfseries 1409} (2014)
  087}, (erratum \href{http://dx.doi.org/10.1007/JHEP10(2014)106}{\textit{ibid}
  \textbf{1410} (2014), 106}),
\href{http://arxiv.org/abs/1408.1682}{{\ttfamily arXiv:1408.1682 [hep-ex]}}.

\bibitem{tthbb}
{ATLAS} Collaboration, {\em {Search for the Standard Model Higgs boson produced
  in association with top quarks and decaying into $b\bar{b}$ in pp collisions
  at $\sqrt{s}$ = 8 TeV with the ATLAS detector}},
  \href{http://dx.doi.org/10.1140/epjc/s10052-015-3543-1}{Eur.~Phys.~J.~C
  {\bfseries 75} (2015) 349}, \href{http://arxiv.org/abs/1503.05066}{{\ttfamily
  arXiv:1503.05066 [hep-ex]}}.

\bibitem{2014tthdiph}
{ATLAS} Collaboration, {\em Search for $H\rightarrow \gamma \gamma$ produced in
  association with top quarks and constraints on the Yukawa coupling between
  the top quark and the Higgs boson using data taken at 7 TeV and 8 TeV with
  the ATLAS detector},
  \href{http://dx.doi.org/10.1016/j.physletb.2014.11.049}{Phys. Lett. B
  {\bfseries 740} (2015) 222}, \href{http://arxiv.org/abs/1409.3122}{{\ttfamily
  arXiv:1409.3122 [hep-ex]}}.

\bibitem{detectorpaper}
{ATLAS} Collaboration, {\em ATLAS Experiment at the CERN Large Hadron
  Collider}, \href{http://dx.doi.org/10.1088/1748-0221/3/08/S08003}{JINST
  {\bfseries 3} (2008) S08003}.

\bibitem{Dawson:2003zu}
S.~Dawson, C.~Jackson, L.~Orr, L.~Reina, and D.~Wackeroth, {\em Associated
  Higgs production with top quarks at the Large Hadron Collider: NLO QCD
  corrections},
  \href{http://dx.doi.org/10.1103/PhysRevD.68.034022}{Phys.~Rev.~D {\bfseries
  68} (2003) 034022},
\href{http://arxiv.org/abs/hep-ph/0305087}{{\ttfamily arXiv:hep-ph/0305087
  [hep-ph]}}.

\bibitem{Reina:2001sf}
L.~Reina and S.~Dawson, {\em Next-to-leading order results for $t\bar{t} h$
  production at the Tevatron},
  \href{http://dx.doi.org/10.1103/PhysRevLett.87.201804}{Phys.~Rev.~Lett.
  {\bfseries 87} (2001) 201804},
\href{http://arxiv.org/abs/hep-ph/0107101}{{\ttfamily arXiv:hep-ph/0107101
  [hep-ph]}}.

\bibitem{Beenakker:2002nc}
W.~Beenakker {et~al.}, {\em NLO QCD corrections to $t\bar{t} H$ production in
  hadron collisions},
  \href{http://dx.doi.org/10.1016/S0550-3213(03)00044-0}{Nucl.~Phys.~B
  {\bfseries 653} (2003) 151},
\href{http://arxiv.org/abs/hep-ph/0211352}{{\ttfamily arXiv:hep-ph/0211352
  [hep-ph]}}.

\bibitem{Beenakker:2001rj}
W.~Beenakker {et~al.}, {\em Higgs radiation off top quarks at the Tevatron and
  the LHC},
  \href{http://dx.doi.org/10.1103/PhysRevLett.87.201805}{Phys.~Rev.~Lett.
  {\bfseries 87} (2001) 201805},
\href{http://arxiv.org/abs/hep-ph/0107081}{{\ttfamily arXiv:hep-ph/0107081
  [hep-ph]}}.

\bibitem{lhcxs}
{LHC Higgs Cross Section Working Group} Collaboration, S.~Dittmaier {et~al.},
  {\em Handbook of LHC Higgs Cross Sections: 1. Inclusive Observables},
  \href{http://arxiv.org/abs/1101.0593}{{\ttfamily arXiv:1101.0593 [hep-ph]}}.

\bibitem{Heinemeyer:2013tqa}
{LHC Higgs Cross Section Working Group} Collaboration, S.~Heinemeyer {et~al.},
  {\em Handbook of LHC Higgs Cross Sections: 3. Higgs Properties},
\href{http://arxiv.org/abs/1307.1347}{{\ttfamily arXiv:1307.1347 [hep-ph]}}.

\bibitem{Aad:2015zhl}
{ATLAS and CMS Collaborations}, {\em {Combined Measurement of the Higgs Boson
  Mass in $pp$ Collisions at $\sqrt{s}=$ 7 and 8 TeV with the ATLAS and CMS
  Experiments}}, \href{http://dx.doi.org/10.1103/PhysRevLett.114.191803}{Phys.
  Rev. Lett. {\bfseries 114} (2015) 191803},
\href{http://arxiv.org/abs/1503.07589}{{\ttfamily arXiv:1503.07589 [hep-ex]}}.

\bibitem{Alwall:2014hca}
J.~Alwall {et~al.}, {\em The automated computation of tree-level and
  next-to-leading order differential cross sections, and their matching to
  parton shower simulations},
  \href{http://dx.doi.org/10.1007/JHEP07(2014)079}{JHEP {\bfseries 1407} (2014)
  079},
\href{http://arxiv.org/abs/1405.0301}{{\ttfamily arXiv:1405.0301 [hep-ph]}}.

\bibitem{Campbell:2012dh}
J.~M. Campbell and R.~K. Ellis, {\em $t \bar{t} W^{\pm}$ production and decay
  at NLO}, \href{http://dx.doi.org/10.1007/JHEP07(2012)052}{JHEP {\bfseries
  1207} (2012) 052},
\href{http://arxiv.org/abs/1204.5678}{{\ttfamily arXiv:1204.5678 [hep-ph]}}.

\bibitem{Garzelli:2012bn}
M.~Garzelli, A.~Kardos, C.~Papadopoulos, and Z.~Trocsanyi, {\em
  $t\bar{t}W^{\pm}$ and $t\bar{t} Z$ Hadroproduction at NLO accuracy in QCD
  with Parton Shower and Hadronization effects},
  \href{http://dx.doi.org/10.1007/JHEP11(2012)056}{JHEP {\bfseries 1211} (2012)
  056},
\href{http://arxiv.org/abs/1208.2665}{{\ttfamily arXiv:1208.2665 [hep-ph]}}.

\bibitem{Campbell:2013yla}
J.~M. Campbell, R.~K. Ellis, and R.~R{\"o}ntsch, {\em Single top production in
  association with a Z boson at the LHC},
  \href{http://dx.doi.org/10.1103/PhysRevD.87.114006}{Phys.~Rev.~D {\bfseries
  87} (2013) 114006},
\href{http://arxiv.org/abs/1302.3856}{{\ttfamily arXiv:1302.3856 [hep-ph]}}.

\bibitem{mg5}
J.~Alwall, M.~Herquet, F.~Maltoni, O.~Mattelaer, and T.~Stelzer, {\em MadGraph
  5 : Going Beyond}, \href{http://dx.doi.org/10.1007/JHEP06(2011)128}{JHEP
  {\bfseries 1106} (2011) 128},
  \href{http://arxiv.org/abs/1106.0522}{{\ttfamily arXiv:1106.0522 [hep-ex]}}.

\bibitem{Campbell:2011bn}
J.~M. Campbell, R.~K. Ellis, and C.~Williams, {\em Vector boson pair production
  at the LHC}, \href{http://dx.doi.org/10.1007/JHEP07(2011)018}{JHEP {\bfseries
  1107} (2011) 018},
\href{http://arxiv.org/abs/1105.0020}{{\ttfamily arXiv:1105.0020 [hep-ph]}}.

\bibitem{Czakon:2011xx}
M.~Czakon and A.~Mitov, {\em {Top++: A Program for the Calculation of the
  Top-Pair Cross-Section at Hadron Colliders}},
  \href{http://dx.doi.org/10.1016/j.cpc.2014.06.021}{Comput. Phys. Commun.
  {\bfseries 185} (2014) 2930},
\href{http://arxiv.org/abs/1112.5675}{{\ttfamily arXiv:1112.5675 [hep-ph]}}.

\bibitem{Kidonakis:2010tc}
N.~Kidonakis, {\em NNLL resummation for s-channel single top quark production},
  \href{http://dx.doi.org/10.1103/PhysRevD.81.054028}{Phys.~Rev.~D {\bfseries
  81} (2010) 054028},
\href{http://arxiv.org/abs/1001.5034}{{\ttfamily arXiv:1001.5034 [hep-ph]}}.

\bibitem{Kidonakis:2011wy}
N.~Kidonakis, {\em Next-to-next-to-leading-order collinear and soft gluon
  corrections for t-channel single top quark production},
  \href{http://dx.doi.org/10.1103/PhysRevD.83.091503}{Phys.~Rev.~D {\bfseries
  83} (2011) 091503},
\href{http://arxiv.org/abs/1103.2792}{{\ttfamily arXiv:1103.2792 [hep-ph]}}.

\bibitem{Kidonakis:2010ux}
N.~Kidonakis, {\em Two-loop soft anomalous dimensions for single top quark
  associated production with a W or H},
  \href{http://dx.doi.org/10.1103/PhysRevD.82.054018}{Phys.~Rev.~D {\bfseries
  82} (2010) 054018},
\href{http://arxiv.org/abs/1005.4451}{{\ttfamily arXiv:1005.4451 [hep-ph]}}.

\bibitem{Martin:2009iq}
A.~Martin, W.~J. Stirling, R.~S. Thorne, and G.~Watt, {\em Parton distributions
  for the LHC},
  \href{http://dx.doi.org/10.1140/epjc/s10052-009-1072-5}{Eur.~Phys.~J.~C
  {\bfseries 63} (2009) 189},
\href{http://arxiv.org/abs/0901.0002}{{\ttfamily arXiv:0901.0002 [hep-ph]}}.

\bibitem{Anastasiou:2003ds}
C.~Anastasiou, L.~J. Dixon, K.~Melnikov, and F.~Petriello, {\em High precision
  QCD at hadron colliders: Electroweak gauge boson rapidity distributions at
  NNLO}, \href{http://dx.doi.org/10.1103/PhysRevD.69.094008}{Phys.~Rev.~D
  {\bfseries 69} (2004) 094008},
\href{http://arxiv.org/abs/hep-ph/0312266}{{\ttfamily arXiv:hep-ph/0312266
  [hep-ph]}}.

\bibitem{Bevilacqua:2011xh}
G.~Bevilacqua {et~al.}, {\em HELAC-NLO},
  \href{http://dx.doi.org/10.1016/j.cpc.2012.10.033}{Comput.~Phys.~Commun.
  {\bfseries 184} (2013) 986},
\href{http://arxiv.org/abs/1110.1499}{{\ttfamily arXiv:1110.1499 [hep-ph]}}.

\bibitem{Garzelli:2011vp}
M.~Garzelli, A.~Kardos, C.~Papadopoulos, and Z.~Trocsanyi, {\em {Standard Model
  Higgs boson production in association with a top anti-top pair at NLO with
  parton showering}},
  \href{http://dx.doi.org/10.1209/0295-5075/96/11001}{Europhys.~Lett.
  {\bfseries 96} (2011) 11001},
\href{http://arxiv.org/abs/1108.0387}{{\ttfamily arXiv:1108.0387 [hep-ph]}}.

\bibitem{Pythia8}
T.~Sj{\"o}strand, S.~Mrenna, and P.~Skands, {\em A brief introduction to PYTHIA
  8.1}, \href{http://dx.doi.org/10.1016/j.cpc.2008.01.036}{Comput. Phys.
  Commun. {\bfseries 178} (2008) 852},
  \href{http://arxiv.org/abs/0710.3820}{{\ttfamily arXiv:0710.3820 [hep-ph]}}.

\bibitem{ct10}
H.-L. Lai {et~al.}, {\em New parton distributions for collider physics},
  \href{http://dx.doi.org/10.1103/PhysRevD.82.074024}{Phys. Rev. D {\bfseries
  82} (2010) 074024}, \href{http://arxiv.org/abs/1007.2241}{{\ttfamily
  arXiv:1007.2241 [hep-ph]}}.

\bibitem{cteq6l1}
J.~Pumplin, D.~R. Stump, J.~Huston, H.-L. Lai, P.~M. Nadolsky, and W.~K. Tung,
  {\em New Generation of Parton Distributions with Uncertainties from Global
  QCD Analysis}, \href{http://dx.doi.org/10.1088/1126-6708/2002/07/012}{JHEP
  {\bfseries 0207} (2012) 012},
\href{http://arxiv.org/abs/hep-ph/0201195}{{\ttfamily arXiv:hep-ph/0201195
  [hep-ph]}}.

\bibitem{cteq6}
P.~M. Nadolsky {et~al.}, {\em Implications of CTEQ global analysis for collider
  observables}, \href{http://dx.doi.org/10.1103/PhysRevD.78.013004}{Phys. Rev.
  D {\bfseries 78} (2008) 013004},
  \href{http://arxiv.org/abs/0802.0007}{{\ttfamily arXiv:0802.0007 [hep-ph]}}.

\bibitem{ATLASUETune0}
{ATLAS} Collaboration, {\em Summary of ATLAS Pythia 8 tunes},
  ATL-PHYS-PUB-2012-003.
  \href{https://cds.cern.ch/record/1474107}{https://cds.cern.ch/record/1474107}.

\bibitem{Nason:2004rx}
P.~Nason, {\em A New method for combining NLO QCD with shower Monte Carlo
  algorithms}, \href{http://dx.doi.org/10.1088/1126-6708/2004/11/040}{JHEP
  {\bfseries 0411} (2004) 040},
\href{http://arxiv.org/abs/hep-ph/0409146}{{\ttfamily arXiv:hep-ph/0409146
  [hep-ph]}}.

\bibitem{Frixione:2007vw}
S.~Frixione, P.~Nason, and C.~Oleari, {\em Matching NLO QCD computations with
  Parton Shower simulations: the POWHEG method},
  \href{http://dx.doi.org/10.1088/1126-6708/2007/11/070}{JHEP {\bfseries 0711}
  (2007) 070},
\href{http://arxiv.org/abs/0709.2092}{{\ttfamily arXiv:0709.2092 [hep-ph]}}.

\bibitem{Alioli:2010xd}
S.~Alioli {et~al.}, {\em A general framework for implementing NLO calculations
  in shower Monte Carlo programs: the POWHEG BOX},
  \href{http://dx.doi.org/10.1007/JHEP06(2010)043}{JHEP {\bfseries 1006} (2010)
  043},
\href{http://arxiv.org/abs/1002.2581}{{\ttfamily arXiv:1002.2581 [hep-ph]}}.

\bibitem{Bahr:2008pv}
M.~Bahr {et~al.}, {\em Herwig++ Physics and Manual},
  \href{http://dx.doi.org/10.1140/epjc/s10052-008-0798-9}{Eur.~Phys.~J.~C
  {\bfseries 58} (2008) 639},
\href{http://arxiv.org/abs/0803.0883}{{\ttfamily arXiv:0803.0883 [hep-ph]}}.

\bibitem{Sherstnev:2007nd}
A.~Sherstnev and R.~Thorne, {\em {Parton Distributions for LO Generators}},
  \href{http://dx.doi.org/10.1140/epjc/s10052-008-0610-x}{Eur.~Phys.~J.~C
  {\bfseries 55} (2008) 553},
\href{http://arxiv.org/abs/0711.2473}{{\ttfamily arXiv:0711.2473 [hep-ph]}}.

\bibitem{Gieseke:2012ft}
S.~Gieseke, C.~Rohr, and A.~Siodmok, {\em {Colour reconnections in Herwig++}},
  \href{http://dx.doi.org/10.1140/epjc/s10052-012-2225-5}{Eur.~Phys.~J.~C
  {\bfseries 72} (2012) 2225},
\href{http://arxiv.org/abs/1206.0041}{{\ttfamily arXiv:1206.0041 [hep-ph]}}.

\bibitem{Pythia6}
T.~Sj{\"o}strand {et~al.}, {\em High-energy-physics event generation with
  Pythia 6.1}, \href{http://dx.doi.org/10.1016/S0010-4655(00)00236-8}{Comput.\
  Phys.\ Commun. {\bfseries 135} (2001) 238},
  \href{http://arxiv.org/abs/hep-ph/0010017}{{\ttfamily arXiv:hep-ph/0010017
  [hep-ph]}}.

\bibitem{ATLASUETune1}
{ATLAS} Collaboration, {\em ATLAS tunes of \textsc{PYTHIA}6 and PYTHIA8 for
  MC11}, ATL-PHYS-PUB-2011-009.
  \href{http://cds.cern.ch/record/1363300}{http://cds.cern.ch/record/1363300}.

\bibitem{sherpa}
T.~Gleisberg {et~al.}, {\em Event generation with SHERPA 1.1},
  \href{http://dx.doi.org/10.1088/1126-6708/2009/02/007}{JHEP {\bfseries 0902}
  (2009) 007},
\href{http://arxiv.org/abs/0811.4622}{{\ttfamily arXiv:0811.4622 [hep-ph]}}.

\bibitem{powhegVV}
T.~Melia, P.~Nason, R.~Rontsch, and G.~Zanderighi, {\em $W^+W^-$, $WZ$ and $ZZ$
  production in the POWHEG BOX},
  \href{http://dx.doi.org/10.1007/JHEP11(2011)078}{JHEP {\bfseries 1111} (2011)
  078}, \href{http://arxiv.org/abs/1107.5051}{{\ttfamily arXiv:1107.5051
  [hep-ph]}}.

\bibitem{Binoth:2008pr}
T.~Binoth, N.~Kauer, and P.~Mertsch, {\em Gluon-induced QCD corrections to $pp
  \to ZZ \to l \bar l l' \bar l'$},
\href{http://arxiv.org/abs/0807.0024}{{\ttfamily arXiv:0807.0024 [hep-ph]}}.

\bibitem{herwig}
G.~Corcella {et~al.}, {\em HERWIG 6.5: an event generator for Hadron Emission
  Reactions With Interfering Gluons (including supersymmetric processes)},
  \href{http://dx.doi.org/10.1088/1126-6708/2001/01/010}{JHEP {\bfseries 0101}
  (2001) 010},
\href{http://arxiv.org/abs/hep-ph/0011363}{{\ttfamily arXiv:hep-ph/0011363
  [hep-ph]}}.

\bibitem{ATLASUETune2}
{ATLAS} Collaboration, {\em New ATLAS event generator tunes to 2010 data},
  ATL-PHYS-PUB-2011-008.
  \href{http://cds.cern.ch/record/1345343}{http://cds.cern.ch/record/1345343}.

\bibitem{powhegtt}
S.~Frixione, G.~Ridolfi, and P.~Nason, {\em A positive-weight
  next-to-leading-order Monte Carlo for heavy flavour hadroproduction},
  \href{http://dx.doi.org/10.1088/1126-6708/2007/09/126}{JHEP {\bfseries 0709}
  (2009) 126}, \href{http://arxiv.org/abs/0707.3088}{{\ttfamily arXiv:0707.3088
  [hep-ph]}}.

\bibitem{perugia}
P.~Skands, {\em Tuning Monte Carlo Generators: The Perugia Tunes},
  \href{http://dx.doi.org/10.1103/PhysRevD.82.074018}{Phys.~Rev.~D {\bfseries
  82} (2010) 074018},
\href{http://arxiv.org/abs/1005.3457}{{\ttfamily arXiv:1005.3457 [hep-ph]}}.

\bibitem{powhegstp}
E.~Re, {\em Single-top Wt-channel production matched with parton showers using
  the POWHEG method},
  \href{http://dx.doi.org/10.1140/epjc/s10052-011-1547-z}{Eur. Phys. J. C
  {\bfseries 71} (2011) 1547}, \href{http://arxiv.org/abs/1009.2450}{{\ttfamily
  arXiv:1009.2450 [hep-ph]}}.

\bibitem{powhegstp2}
S.~Alioli, P.~Nason, C.~Oleari, and E.~Re, {\em NLO single-top production
  matched with shower in POWHEG: s- and t-channel contributions},
  \href{http://dx.doi.org/10.1088/1126-6708/2009/09/111}{JHEP {\bfseries 0909}
  (2009) 111}, (erratum
  \href{http://dx.doi.org/10.1007/JHEP02(2010)011}{\textit{ibid} \textbf{1002}
  (2010), 011}), \href{http://arxiv.org/abs/0907.4076}{{\ttfamily
  arXiv:0907.4076 [hep-ph]}}.

\bibitem{alpgen}
M.~L. Mangano {et~al.}, {\em ALPGEN, a generator for hard multiparton processes
  in hadronic collisions},
  \href{http://dx.doi.org/10.1088/1126-6708/2003/07/001}{JHEP {\bfseries 0307}
  (2003) 001}, \href{http://arxiv.org/abs/hep-ph/0206293}{{\ttfamily
  arXiv:hep-ph/0206293 [hep-ph]}}.

\bibitem{Djouadi:1997yw}
A.~Djouadi, J.~Kalinowski, and M.~Spira, {\em HDECAY: A Program for Higgs boson
  decays in the standard model and its supersymmetric extension},
  \href{http://dx.doi.org/10.1016/S0010-4655(97)00123-9}{Comput.~Phys.~Commun.
  {\bfseries 108} (1998) 56},
\href{http://arxiv.org/abs/hep-ph/9704448}{{\ttfamily arXiv:hep-ph/9704448
  [hep-ph]}}.

\bibitem{Bredenstein:2006rh}
A.~Bredenstein, A.~Denner, S.~Dittmaier, and M.~Weber, {\em Precise predictions
  for the Higgs-boson decay $H \to WW/ZZ \to$ 4 leptons},
  \href{http://dx.doi.org/10.1103/PhysRevD.74.013004}{Phys.~Rev.~D {\bfseries
  74} (2006) 013004},
\href{http://arxiv.org/abs/hep-ph/0604011}{{\ttfamily arXiv:hep-ph/0604011
  [hep-ph]}}.

\bibitem{Actis:2008ts}
S.~Actis, G.~Passarino, C.~Sturm, and S.~Uccirati, {\em NNLO Computational
  Techniques: The Cases $H \to \gamma\gamma$ and $H \to g g$},
  \href{http://dx.doi.org/10.1016/j.nuclphysb.2008.11.024}{Nucl.~Phys.~B
  {\bfseries 811} (2009) 182},
\href{http://arxiv.org/abs/0809.3667}{{\ttfamily arXiv:0809.3667 [hep-ph]}}.

\bibitem{Denner:2011mq}
A.~Denner, S.~Heinemeyer, I.~Puljak, D.~Rebuzzi, and M.~Spira, {\em Standard
  Model Higgs-Boson Branching Ratios with Uncertainties},
  \href{http://dx.doi.org/10.1140/epjc/s10052-011-1753-8}{Eur.~Phys.~J.~C
  {\bfseries 71} (2011) 1753},
\href{http://arxiv.org/abs/1107.5909}{{\ttfamily arXiv:1107.5909 [hep-ph]}}.

\bibitem{topdiff_7TEV}
{ATLAS} Collaboration, {\em Measurements of normalized differential cross
  sections for $t\bar{t}$ production in pp collisions at $\sqrt{s}=$ 7 TeV
  using the ATLAS detector},
  \href{http://dx.doi.org/10.1103/PhysRevD.90.072004}{Phys.~Rev.~D {\bfseries
  90} (2014) 072004},
\href{http://arxiv.org/abs/1407.0371}{{\ttfamily arXiv:1407.0371 [hep-ex]}}.

\bibitem{Frixione:2008yi}
S.~Frixione, E.~Laenen, P.~Motylinski, B.~R. Webber, and C.~D. White, {\em
  {Single-top hadroproduction in association with a W boson}},
  \href{http://dx.doi.org/10.1088/1126-6708/2008/07/029}{JHEP {\bfseries 0807}
  (2008) 029},
\href{http://arxiv.org/abs/0805.3067}{{\ttfamily arXiv:0805.3067 [hep-ph]}}.

\bibitem{mlm}
M.~L. Mangano {et~al.}, {\em Multijet matrix elements and shower evolution in
  hadronic collisions: $Wb\bar{b}+n$ jets as a case study},
  \href{http://dx.doi.org/(10.1016/S0550-3213(02)00249-3)}{Nucl.\ Phys.
  {\bfseries B 632} (2002) 343},
  \href{http://arxiv.org/abs/hep-ph/0108069}{{\ttfamily arXiv:hep-ph/0108069
  [hep-ph]}}.

\bibitem{PhotosPaper}
P.~Golonka and Z.~W{\c{a}}s, {\em PHOTOS Monte Carlo: a precision tool for QED
  corrections in $Z$ and $W$ decays},
  \href{http://dx.doi.org/10.1140/epjc/s2005-02396-4}{Eur. Phys. J C {\bfseries
  45} (2006) 97}, \href{http://arxiv.org/abs/hep-ph/0506026}{{\ttfamily
  arXiv:hep-ph/0506026 [hep-ph]}}.

\bibitem{TauolaPaper}
J.~Stanis{\l}aw, J.~K{\"u}hn, and Z.~W{\c{a}}s, {\em TAUOLA - a library of
  Monte Carlo programs to simulate decays of polarized $\tau$ leptons},
  \href{http://dx.doi.org/10.1016/0010-4655(91)90038-M}{Comput. Phys. Commun.
  {\bfseries 64} (1991) 275}.

\bibitem{Butterworth:1996zw}
J.~Butterworth, J.~R. Forshaw, and M.~Seymour, {\em {Multiparton interactions
  in photoproduction at HERA}},
  \href{http://dx.doi.org/10.1007/s002880050286}{Z.~Phys.~C {\bfseries 72}
  (1996) 637},
\href{http://arxiv.org/abs/hep-ph/9601371}{{\ttfamily arXiv:hep-ph/9601371
  [hep-ph]}}.

\bibitem{geant4}
S.~Agostinelli {et~al.}, {\em Geant4: a simulation toolkit},
  \href{http://dx.doi.org/10.1016/S0168-9002(03)01368-8}{Nucl. Instrum. Meth.
  {\bfseries A~506} (2003) 250}.

\bibitem{atlasSim}
{ATLAS} Collaboration, {\em The ATLAS Simulation Infrastructure},
  \href{http://dx.doi.org/10.1140/epjc/s10052-010-1429-9}{Eur.~Phys.~J.~C
  {\bfseries 70} (2010) 823}, \href{http://arxiv.org/abs/1005.4568}{{\ttfamily
  arXiv:1005.4568 [physics.ins-det]}}.

\bibitem{ATLAS:1300517}
{ATLAS} Collaboration, {\em The simulation principle and performance of the
  ATLAS fast calorimeter simulation FastCaloSim}, ATL-PHYS-PUB-2010-013.
  \href{http://cds.cern.ch/record/1300517}{http://cds.cern.ch/record/1300517}.

\bibitem{ATLASUETune3}
{ATLAS} Collaboration, {\em Further ATLAS tunes of \textsc{PYTHIA}6 and Pythia
  8}, ATL-PHYS-PUB-2011-014.
  \href{https://cds.cern.ch/record/1400677}{https://cds.cern.ch/record/1400677}.

\bibitem{Aad:2014fxa}
{ATLAS} Collaboration, {\em Electron reconstruction and identification
  efficiency measurements with the ATLAS detector using the 2011 LHC
  proton-proton collision data},
  \href{http://dx.doi.org/10.1140/epjc/s10052-014-2941-0}{Eur.~Phys.~J.~C
  {\bfseries 74} (2014) 2941},
\href{http://arxiv.org/abs/1404.2240}{{\ttfamily arXiv:1404.2240 [hep-ex]}}.

\bibitem{ATLAS-CONF-2014-032}
{ATLAS} Collaboration, {\em Electron efficiency measurements with the ATLAS
  detector using the 2012 LHC proton-proton collision data},
  ATLAS-CONF-2014-032.
  \href{http://cds.cern.ch/record/1706245}{http://cds.cern.ch/record/1706245}.

\bibitem{mureco}
{ATLAS} Collaboration, {\em Measurement of the muon reconstruction performance
  of the ATLAS detector using 2011 and 2012 LHC proton-proton collision data},
  \href{http://dx.doi.org/10.1140/epjc/s10052-014-3130-x}{Eur.~Phys.~J.~C
  {\bfseries 74} (2014) 3130}, \href{http://arxiv.org/abs/1407.3935}{{\ttfamily
  arXiv:1407.3935 [hep-ex]}}.

\bibitem{ATLASTAUIDnew}
{ATLAS} Collaboration, {\em Identification and energy calibration of
  hadronically decaying tau leptons with the ATLAS experiment in pp collisions
  at $\sqrt{s}$ = 8 TeV},
  \href{http://dx.doi.org/10.1140/epjc/s10052-015-3500-z}{Eur.\ Phys.\ J.\ C
  {\bfseries 75} (2015) 303}, \href{http://arxiv.org/abs/1412.7086}{{\ttfamily
  arXiv:1412.7086 [hep-ph]}}.

\bibitem{ref:Cacciari2008}
M.~Cacciari, G.~P. Salam, and G.~Soyez, {\em The anti-$k_{t}$ jet clustering
  algorithm}, \href{http://dx.doi.org/10.1088/1126-6708/2008/04/063}{JHEP
  {\bfseries 0804} (2008) 063},
  \href{http://arxiv.org/abs/0802.1189}{{\ttfamily arXiv:0802.1189 [hep-ph]}}.

\bibitem{ref:Cacciari2006}
M.~Cacciari and G.~P. Salam, {\em Dispelling the $N^3$ myth for the $k_t$
  jet-finder}, \href{http://dx.doi.org/10.1016/j.physletb.2006.08.037}{Phys.
  Lett. B {\bfseries 641} (2006) 57},
  \href{http://arxiv.org/abs/hep-ph/0512210}{{\ttfamily arXiv:hep-ph/0512210
  [hep-ph]}}.

\bibitem{ref:fastjet}
M.~Cacciari, G.~P. Salam, and G.~Soyez, {\em FastJet User Manual},
  \href{http://dx.doi.org/10.1140/epjc/s10052-012-1896-2}{Eur.~Phys.~J.~C
  {\bfseries 72} (2012) 1896}, \href{http://arxiv.org/abs/1111.6097}{{\ttfamily
  arXiv:1111.6097 [hep-ph]}}.

\bibitem{LCW1}
C.~Cojocaru {et~al.}, {\em Hadronic calibration of the ATLAS liquid argon
  end-cap calorimeter in the pseudorapidity region $1.6<|\eta|<1.8$ in beam
  tests}, \href{http://dx.doi.org/10.1016/j.nima.2004.05.133}{Nucl. Instrum.
  Meth. {\bfseries A~531} (2004) 481},
  \href{http://arxiv.org/abs/physics/0407009}{{\ttfamily arXiv:physics/0407009
  [physics]}}.

\bibitem{LCW2}
T.~Barillari {et~al.}, {\em Local hadronic calibration}, ATL-LARG-PUB-2009-001.
  \href{http://cds.cern.ch/record/1112035}{http://cds.cern.ch/record/1112035}.

\bibitem{JES}
{ATLAS} Collaboration, {\em Jet energy measurement with the ATLAS detector in
  proton-proton collisions at $\sqrt{s}=$ 7 TeV},
  \href{http://dx.doi.org/10.1140/epjc/s10052-013-2304-2}{Eur.~Phys.~J.~C
  {\bfseries 73} (2013) 2304},
\href{http://arxiv.org/abs/1112.6426}{{\ttfamily arXiv:1112.6426 [hep-ex]}}.

\bibitem{JER}
{ATLAS} Collaboration, {\em Jet energy resolution in proton-proton collisions
  at $\sqrt{s}=$ 7 TeV recorded in 2010 with the ATLAS detector},
  \href{http://dx.doi.org/10.1140/epjc/s10052-013-2306-0}{Eur.~Phys.~J.~C
  {\bfseries 73} (2013) 2306},
\href{http://arxiv.org/abs/1210.6210}{{\ttfamily arXiv:1210.6210 [hep-ex]}}.

\bibitem{ATLAS-CONF-2014-046}
{ATLAS} Collaboration, {\em Calibration of the performance of $b$-tagging for
  $c$ and light-flavour jets in the 2012 ATLAS data}, ATLAS-CONF-2014-046.
  \href{http://cds.cern.ch/record/1741020}{http://cds.cern.ch/record/1741020}.

\bibitem{ATLAS-CONF-2014-004}
{ATLAS} Collaboration, {\em Calibration of $b$-tagging using dileptonic top
  pair events in a combinatorial likelihood approach with the ATLAS
  experiment}, ATLAS-CONF-2014-004.
  \href{http://cds.cern.ch/record/1664335}{http://cds.cern.ch/record/1664335}.

\bibitem{openloops}
F.~Cascioli, P.~Maierh{\"o}fer, and S.~Pozzorini, {\em Scattering Amplitudes
  with Open Loops},
  \href{http://dx.doi.org/10.1103/PhysRevLett.108.111601}{Phys.\ Rev.\ Lett.
  {\bfseries 108} (2012) 111601},
  \href{http://arxiv.org/abs/1111.5206}{{\ttfamily arXiv:1111.5206 [hep-ph]}}.

\bibitem{Aad:2013ucp}
{ATLAS} Collaboration, {\em Improved luminosity determination in pp collisions
  at $\sqrt{s} = $ 7~TeV using the ATLAS detector at the LHC},
  \href{http://dx.doi.org/10.1140/epjc/s10052-013-2518-3}{Eur.~Phys.~J.~C
  {\bfseries 73} (2013) 2518},
\href{http://arxiv.org/abs/1302.4393}{{\ttfamily arXiv:1302.4393 [hep-ex]}}.

\bibitem{asym}
G.~Cowan, K.~Cranmer, E.~Gross, and O.~Vitells, {\em Asymptotic formulae for
  likelihood-based tests of new physics},
  \href{http://dx.doi.org/10.1140/epjc/s10052-011-1554-0}{Eur.~Phys.~J.~C
  {\bfseries 71} (2011) 1554}, (erratum
  \href{10.1140/epjc/s10052-013-2501-z}{\textit{ibid} \textbf{73} (2013),
  2501}), \href{http://arxiv.org/abs/1007.1727}{{\ttfamily arXiv:1007.1727
  [physics.data-an]}}.

\bibitem{cls}
A.~L. Read, {\em Presentation of search results: The CL(s) technique},
  \href{http://dx.doi.org/10.1088/0954-3899/28/10/313}{J.~Phys.~G {\bfseries
  28} (2002) 2693}.

\bibitem{ATLAScouplings}
{ATLAS} Collaboration, {\em {Measurements of the Higgs boson production and
  decay rates and coupling strengths using $pp$ collision data at $\sqrt{s}=$ 7
  and 8 TeV in the ATLAS experiment}}, submitted to Eur.\ Phys.\ J.\ C (2015),
\href{http://arxiv.org/abs/1507.04548}{{\ttfamily arXiv:1507.04548 [hep-ex]}}.

\end{thebibliography}\endgroup
